\documentclass[lettersize,journal]{IEEEtran}
\usepackage{amsmath,amsfonts}
\usepackage{algorithmic}
\usepackage{algorithm}
\usepackage{array}
\usepackage{textcomp}
\usepackage{stfloats}
\usepackage{url}
\usepackage{verbatim}
\usepackage{graphicx}
\usepackage{cite}

\usepackage{xcolor,colortbl}
\usepackage{tabularray}
\newcommand{\hcor}[1]{\textcolor[rgb]{1,0,0}{#1}}
\usepackage{svg}
\usepackage{nicematrix, subcaption}

\definecolor{Gray}{gray}{0.85}
\definecolor{LightCyan}{rgb}{0.88,1,1}
\newcolumntype{a}{>{\columncolor{Gray}}c}
\newcolumntype{f}{>{\columncolor{LightCyan}}c}

\begin{document}

\title{Human Body Segment Volume Estimation with Two RGB-D Cameras.}

\author{Giulia Bassani, Emilio Maoddi, Usman Asghar, Carlo Alberto Avizzano, and Alessandro Filippeschi
\thanks{''This work was supported by the BRIEF “Biorobotics Research and Innovation Engineering Facilities” project (Project identification code IR0000036) funded under the National Recovery and Resilience Plan (NRRP), Mission 4 Component 2 Investment 3.1 of Italian Ministry of University and Research funded by the European Union – NextGenerationEU.''}
\thanks{G. Bassani and U. Asghar are with the Institute of Mechanical Intelligence, Scuola Superiore Sant'Anna, 56124, Pisa, Italy (e-mail: giulia.bassani@santannapisa.it)}
\thanks{E. Maoddi is with Leonardo Innovation Labs, Leonardo S.p.A., 21017 Cascina Costa di Samarate (VA), Italy (e-mail: emilio.maoddi@leonardo.com)}
\thanks{C. A. Avizzano and A. Filippeschi are with the Institute of Mechanical Intelligence and the Department of Excellence in Robotics and AI, Scuola Superiore Sant'Anna, 56124, Pisa, Italy (e-mail: carloalberto.avizzano@santannapisa.it and alessandro.filippeschi@santannapisa.it)}}

\markboth{Journal of \LaTeX\ Class Files,~Vol.~14, No.~8, August~2021}%
{Shell \MakeLowercase{\textit{et al.}}: A Sample Article Using IEEEtran.cls for IEEE Journals}


\maketitle

\begin{abstract}
In the field of human biometry, accurately estimating the volume of the whole body and its individual segments is of fundamental importance. Such measurements support a wide range of applications that include assessing health, optimizing ergonomic design, and customizing biomechanical models.
In this work, we presented a Body Segment Volume Estimation (BSV) system to automatically compute whole-body and segment volumes using only two RGB-D cameras, thus limiting the system complexity.
However, to maintain the accuracy comparable to 3D laser scanners, we enhanced the As-Rigid-As-Possible (ARAP) non-rigid registration techniques, disconnecting its energy from the single triangle mesh. Thus, we improved the geometrical coherence of the reconstructed mesh, especially in the lateral gap areas.
We evaluated BSV starting from the RGB-D camera performances, through the results obtained with FAUST dataset human body models, and comparing with a state-of-the-art work, up to real acquisitions. It showed superior ability in accurately estimating human body volumes, and it allows evaluating volume ratios between proximal and distal body segments, which are useful indexes in many clinical applications.
\end{abstract}

\begin{IEEEkeywords}
Non-rigid registration, ARAP, Segmentation, 3D Reconstruction, Depth
\end{IEEEkeywords}

\section{Introduction} \label{sec:Intro}

\IEEEPARstart{I}{n} the field of human biometry, the estimation of whole body volume and of its parts plays an essential role in many fields such as health, ergonomics, and sport. Estimating the volume of the body's segments provides an indirect insight into the distribution of mass along the body, which enhances the information provided by the simple measurement of the height and the weight of a person.

Volume distribution is closely linked to an individual's health and medical status. For instance, the Body Volume Index (BVI) has been proposed as an advanced tool for assessing body shape and weight distribution that, differently from the Body Mass Index (BMI) \cite{keys2014indices}, offers a detailed view of how fat and muscles are distributed across the body. Indeed, epidemiological data showed that, at any BMI level, the central distribution of adiposity increases risks for diseases correlated with overweight and obesity \cite{piche2018overview}.
The estimation of body mass distribution has a clear application in the biomechanics of sport, for both the analysis of performance \cite{payton2007biomechanical} and the prevention of injuries \cite{lloyd2024future}. Personalized biomechanical models based on  \textit{in-vivo} estimation of the body segments' volume and mass are fundamental for equipment design \cite{stefanyshyn2015biomechanics} and ergonomic assessment, even beyond the sport domain \cite{stevenson2004suite}.

The historical "golden standard" for estimating body volume is UnderWater Weighing (UWW) \cite{katch1967estimation}, which provides an accurate measurement regardless of the complexity of the shape of the submerged object. However, the procedure is long, requires expensive equipment, the measuring apparatus is cumbersome, and is limited to static acquisitions.
Imaging techniques, such as computed tomography and magnetic resonance imaging, can generate highly detailed images to estimate volumes and also interpret the regional fat distribution, distinguishing between subcutaneous fat and visceral adipose tissue, which is an even better predictor of health risks \cite{seidell1990imaging}. However, their usage is limited mainly by the high costs.

In contrast, infrared and visible light technologies are a viable way to obtain devices that can accurately reconstruct and estimate body volume, and at the same time be affordable for a large diffusion both in the healthcare systems and in sports facilities. 
3D full-body surface scanners, typically based on laser scanning or pattern light projection, provide a less time-consuming and invasive way to measure a person's body shape and volumes. They are composed of many cameras placed at fixed locations around the subject and have been commercialized since the 1990s. Thus, it has been largely proven that they have comparable accuracy to the traditional UWW technique \cite{bartol2021review}. However, their cost and encumbrance strongly limit their use \cite{daanen20133d}.
%Thus, it has been largely proven that they have higher repeatability and velocity than human expert measurers and comparable accuracy to the traditional UWW technique \cite{bartol2021review}. However, their cost and encumbrance strongly limit their use \cite{daanen20133d}.
Simpler vision-based systems include RGB and depth cameras. Although RGB cameras have been used with Deep-Learning (DL) algorithms to reconstruct the whole human body, their lack of depth information limits their accuracy and hinders their applications in health and sports biomechanics. 
Instead, depth-camera-based reconstruction has greater accuracy, using knowledge of the depth. RGB-D cameras, such as Microsoft Kinect and Realsense L515, based on structured light scanner or Laser Imaging Detection and Ranging (LiDAR) technologies, respectively, provide fast depth map creation with aligned color information at low cost and size. In addition, especially in healthcare, the possibility of texturizing the acquired Point Cloud (PC) gives useful insights into the patient's health that can help the doctor in the diagnosis.
%\hcor{3D scanners, composed of many RGB-D cameras placed at fixed locations around the subject, have comparable accuracy to the traditional UWW technique \cite{bartol2021review}. However, their cost and encumbrance strongly limit their use \cite{daanen20133d}. Thus, a low-cost vision-based approach that employs a limited number of sensors while maintaining high speed and accuracy needs to be investigated.}

3-D body model reconstruction with depth cameras targets the reconstruction of a regular body shape surface from a PC, and when using a limited number of depth cameras, the reconstruction method must ensure geometric coherence, especially in the PC gap area caused by the lack of overlapping between camera acquisitions.
%A limited number of depth cameras prevents the acquisitions from overlapping, causing gaps in the obtained PC. These gaps may cause irregularities in the registration or make it impossible. Therefore, in these cases a 3D body model reconstruction method must ensure geometric coherence, especially in the gap area.
%\hcor{Reconstruction} with depth cameras targets the reconstruction of a regular body shape surface from a PC (i.e., registration).
Approaches to solve this problem include parametric and non-parametric methods. The former involve parametric human body models, e.g., SMPL \cite{loper2023smpl} or SCAPE \cite{anguelov2005scape}, that regress a low-dimensional parameter space generally including shape and pose parameters. 
Researchers largely use parametric reconstruction methods because they provide a fast and computationally efficient way to generate 3D human body models. However, they often rely on simplified representations, where details can be lost, resulting in models with low fidelity, which is insufficient for clinical or biomechanical applications. The non-parametric methods, providing a high-dimensional human body mesh representation able to model the unique characteristics of a subject, are hence preferred.

In this paper, we solve the trade-off between complexity and accuracy by proposing a complete and automatic Body-Segment-Volume (BSV) estimation pipeline to evaluate human full-body and segment volumes using only two RGB-D cameras, employing a non-rigid registration technique able to direct the optimization towards the correct global minimum thanks to a new optimization strategy. The hardware simplicity makes the BSV system affordable for large scale diffusion, such as in general practitioner ambulatories. At the same time, we claim superior accuracy with respect to systems that use at most two cameras. The main contributions of the paper are the BSV system, which includes the hardware setup and processing pipeline; an improvement of the existing non-rigid registration technique; and a comparison of the proposed method against the State-of-the-Art (SoA).

The paper is organized as follows: Section \ref{sec:Back} presents the background, including the presentations of the 3D scanning technologies, the 3D registration methods, and the segment volume estimation methods; Section \ref{sec:BSV} presents BSV detailing all the steps from the system setup to the volume evaluation; Section \ref{sec:Validation} presents the five consecutive validation steps to attest the BSV validity and accuracy; Section \ref{sec:Results} present the results of the validation; Section \ref{sec:Discussion} discusses the results; and finally Section \ref{sec:Conclusion} closes the paper.

\section{Background} \label{sec:Back}

\subsection{3D Scanning Technologies} \label{subsec:3DScanTech}

Stationary scanners generally use either Passive Stereo (PS) or Structured Light (SL) technologies. They are more accurate and reliable and leading many researchers to use them to create 3D human body surface datasets, such as CAESAR \cite{caesar2002} and FAUST \cite{bogo2014faust}, which have been then used as ground truth to validate new anthropometric measurement pipelines and 3D body shape reconstruction algorithms, respectively.
However, many human-centered applications, such as primary health care, need lighter and portable scanners to estimate human body volumes.
To this end, RGB-D cameras enable quick generation of depth maps with synchronized color data, offering a compact and cost-effective solution.
Many researchers used Kinect cameras to reconstruct 3D human bodies \cite{kwok2014volumetric, cui2013kinectavatar, tong2012scanning}. %but the low resolution of the depth maps limited their results 
Cui et al. \cite{cui2013kinectavatar} developed a scanning system based on a single depth camera. They acquired different sets of frames during the subject rotation, ensuring to have enough overlapping areas. However, they need to face both the interferences in the overlapping areas and the inevitable subject movements.
Tong et al. \cite{tong2012scanning} used multiple Kinects to create a 3D full human body model to avoid interferences and reduce misalignment between the different acquisitions. They used three cameras without overlapping regions, two in front of the subject for the upper and lower parts of the body, and the third in the back for the middle part.
Kwok et al. \cite{kwok2014volumetric} used two Kinects, one in front and one in the back, to reconstruct the 3D human body model from incomplete data. To verify the reconstruction quality, they successively acquire depth information both with the RGB-D cameras viewing the central part of the body, from neck to thigh, and an SL full body scanner, concluding that even if their 3D human body model is reliable and usable for manufacturing application, its accuracy is strongly limited by the Kinect low resolution (640 x 480 pixels).
LiDAR RGB-D cameras, which provide more detailed depth maps and are commonly used for topographic mapping, environmental monitoring, and autonomous vehicles \cite{cheng2018registration}, have yet to be investigated in human body shape reconstruction. 
%Lately some researchers employ them also in 3D human body model reconstruction.
To the best of our knowledge, no researchers have employed LiDAR RGB-D devices to estimate human body volumes. Wang et al. \cite{wang2024real} combined LiDAR and RGB data to estimate the person's height.
%anthropometric measurements. They implemented a height measurement system that combines LiDAR with a camera. They used the LiDAR to find the lowest point on the ground and the camera to determine the 2D coordinates of the human face to estimate a person's height. 
Instead, Oberhofer et al. \cite{oberhofer2024feasibility} used the LiDAR sensor included in the iPhone 12 to assess the feasibility of extracting thigh and shank length measurements, concluding that LiDAR technology is promising for contactless anthropometric assessment.

\subsection{3D registration methods} \label{seq:Regmethods}

Using a limited number of RGB-D devices inevitably causes a lack of overlap between different viewpoints. On the one hand, this reduces signal interference in overlapping areas; on the other hand, it presents a challenge in aligning partial scans \cite{su2020robustfusion}.
Template-based methods avoid these problems by warping a 3D full-body high-detailed template mesh to the incomplete PC, thus allowing for filling gaps \cite{liu2016template}. 
%These optimization processes are called mesh registrations, and they can be classified into rigid and non-rigid techniques. 
These optimization processes are called mesh registrations, and when dealing with human bodies that deform non-rigidly due to underlying articulations, non-rigid techniques are used. 
%The former imposes a single transformation on the entire mesh, such as the rigid ICP algorithm \cite{besl1992method}.
%%The latter allows mesh regions to deform differently, modeling non-rigid behaviours due, for instance, to the underlying articulations, and are therefore more promising for the accuracy of the reconstruction. 
%The latter allows mesh regions to deform differently, modeling non-rigid behaviours due, for instance, to the underlying articulations. Thus, determining these deformations is a key challenge in the acquisition and analysis of human bodies that deform non-rigidly. 
% So far, different representations of the deformation field have been used, and the proper one needs to be chosen considering the trade-off between precision and computational cost. 

Non-rigid registration methods \cite{deng2022survey} allow mesh regions to deform differently and share three key components: the transformation that links the two datasets, the similarity metric that assesses their resemblance, and the optimization method that identifies the best transformation parameters and minimizes the error of the similarity metric. The objective function (Eq. \ref{ObjFunc}) typically computes a transformation energy $E$ which is obtained as:   
\begin{equation} \label{ObjFunc}
E = E_{fit} + \alpha E_{reg}.
\end{equation}
where the fitting term $E_{fit}$ decreases as the template model aligns more with the measured PC, and $E_{reg}$ is a regularization term that prevents unrealistic deformations. $\alpha$ is the weight that balances these two terms.
The main differences between approaches lie in how these two components are defined and calculated. The fitting term has been generally represented by \textit{point-to-point} \cite{amberg2007optimal}, \textit{point-to-plane} \cite{li2018articulatedfusion} distances, or combinations thereof \cite{su2020robustfusion}. The regularization term can be a weighted combination of several components, each imposing a distinct constraint on the deformation field. The most common requirements are: the \textit{smoothness} to prevent unrealistic deformed shapes \cite{allen2003space}, the \textit{positional constraints} to ensure that some points stay close to a reference position \cite{allen2003space}, and the \textit{local shape preservation} to preserve the surface locally \cite{sussmuth2008reconstructing}. 
The latter can be expressed with different types of regularization terms. Many researchers imposed the deformation to be locally rigid, i.e., the entire surface undergoes a non-rigid deformation to align with the target, whereas each region locally experiences a nearly rigid transformation. 
%Sometimes, the regularization term is aimed at guaranteeing that affine transformations are close to rigid transformations \cite{dou2016fusion4d, yang2019global}. 
Often, the distance metric is maintained locally by penalizing any changes in the distance between each point and its neighbors using the As-Rigid-As-Possible (ARAP) approach \cite{sorkine2007rigid}. 
%However, since local rigidity must also ensure that the distance metric is maintained locally, some methods enforce this by penalizing any changes in the distance between each point and its neighbors using the As-Rigid-As-Possible (ARAP) approach \cite{sorkine2007rigid}. %, which reduces surface deformation energy (Eq. \ref{ARAP}) to achieve mesh edits that preserve detail.
Yang et al. \cite{yang2019global} employed an ARAP constraint in a sparse non-rigid registration framework to reduce the inward shrinkage of the deformed models, especially when overlapping regions of neighboring scans are small. ARAP allows for preserving the lengths of all the edges as much as possible before and after transformations.
% In addition, since the local rigidity is evaluated based on the edges of each cell, the ARAP approach can be easily extended to volumetric cells, such as tetrahedra. 
% Indeed, Kwok et al. \cite{kwok2014volumetric} employed a volumetric mesh in template fitting to prevent the minimum multiplicity of the surface ARAP, which can lead to an abnormal fitting result.
%the optimization could be applied to a coarse volumetric grid which controls the shape embedded in it, rather than directly to the discrete surface or volume. (vedi ARAP paper)
% And Innmann et al. \cite{innmann2016volumedeform} used an ARAP regularization term to make the highly underconstrained non-rigid tracking problems well posed since they represented the space deformation field with a fine volumetric grid.

%Süßmuth et al. \cite{sussmuth2008reconstructing}

The ARAP algorithm is simple and simultaneously efficient because each optimization step is conceptually similar to Laplacian modeling with a system matrix that needs to be factorized just once and is constant throughout the iterations. 
Therefore, the ARAP algorithm is widely used as both a regularization term and the main fitting term, and different variants have been proposed to improve consistency, especially in the case of large rotations. 
Chen et al. \cite{chen2017rigidity} proposed to use wider local neighborhoods to increase the uniformity of nearby rigid transformations, compared to the classic ARAP, which optimizes rigid transformations in the 1-ring neighborhoods. 
Jiang et al. \cite{jiang2017consistent} employed spokes and rims discrete cells, and introduced a dining term to obtain a consistent ASAP approach to address large deformations.
%Yamazaki et al. \cite{yamazaki2013non} extended the ARAP to the As-Similar-As-Possible (ASAP) approach without applying the linear approximation of the similarity matrix that makes the deformed Laplacian coordinates consistent in ARAP. 
In this work, we used the ARAP algorithm as the main fitting term using a cotangent weighting factor to reduce the asymmetric deformations, and we employed a regularization term to release triangles from shearing and scaling as it would happen if only rotations and translations are allowed. Differently from SoA approaches, we consider the total mesh area in the regularization term, making its energy unconnected to the single triangle mesh areas, which could bring the regularization energy to collapse (Section \ref{seq:Registration}).

\subsection{Segments Volume Estimation Methods} \label{seq:VolumeEstimate}

% Anthropometry is a quantitative discipline dedicated to studying the physical dimensions and structure of the human body that are crucial in different fields (e.g. ergonomics, sports sciences, engineering, and healthcare). 
% Anthropometric measurements collect statistical data regarding the human body's dimensions and physical characteristics such as lengths, circumferences, and volumes. 
% Most of the State-of-the-Art (SoA) works focus on the development of automatic anthropometric measurement systems to estimate lengths and circumferences \hcor{\cite{}}. 
% However, when it comes to body volume, there are still few studies addressing it, and even fewer that consider the volumes of different body segments, even if they are closely linked to an individual's health and medical status.

In the early 2000s, the first works on body volume estimation focused solely on the whole-body volume.
% Solo volume totale
Wells et al. \cite{wells2000assessment} assessed the potential of 3D photonic scanning by comparing the resulting full-body volume estimation with traditional UWW and full-body air displacement plethysmography approaches. 
% Wang et al. \cite{wang2006validation} employed the same scanner, but added a comparison between its performances with dressed and undressed subjects, and estimated part volumes of a mannequin.
%However, they do not estimate body part volumes.
However, they did not estimate body part volumes. % and used a bulky full-body scanner.
% Anche volume segment
Chiu et al. \cite{chiu2018automated} developed a software to estimate both the whole-body volume and the segmental volumes (head, torso, arms, and legs) by applying the Stitched Puppet template matching techniques and evaluated its reliability by comparing it to a manual post-processing technique. However, they also employed a full-body 3D scanner.
% Some researchers aimed at the estimation of human body volume in clinical environments where the subjects lye on a bed and thus the assessment is based solely on the visible frontal view.
More recently, some works targeted the estimation of the volume of body parts.
Pirker et al. \cite{pirker2009human} developed a custom-built examination coach surrounded by 16 stereo cameras and projectors for the illumination to estimate segment volumes (torso, arms, and legs) in a clinical environment.  
Pfitzner et al. \cite{pfitzner2015libra3d} proposed a similar approach, but using only a Kinect placed on the ceiling to allow physicians to treat the patient without hindrances. However, their final aim was to estimate the body weight, and they did not present any results regarding the body volume.
Nuzzi et al. \cite{nuzzi2025measurement} proposed a method to estimate various anthropometric measurements, including the volumes of body segments computed using a 3D Monte Carlo procedure, but they compared their results to anthropometric tables and a model based on truncated cone approximations, which cannot be considered as valid gold standard references.

In 2013, Cook et al. \cite{cook2013using}, aiming at a better normalization of radiation dose, were one of the first research groups to employ a Kinect camera to estimate the volume of the entire body. They segmented and isolated the depth map of the subject from the background, converted it into a PC, and estimated the volume using a convex hull algorithm. However, they doubled the anterior data to obtain the posterior view, and, as they stated, they could have obtained better results using two depth cameras.
% CAMBIA to enable better normalization of dose estimates, and promote more patient-specific protocoling of future CT examinations.
He et al. \cite{he2018volumeter} estimated the whole body volume using a single Kinect camera, which acquires many images during the subject rotation in front of it. They proposed a model–model objective function based on ICP and non-rigid registration to align the different views, and they calculated the volume and other body parameters with truncated signed distance function values of voxels.
However, they did not estimate the different volumes of body segments.
In 2023, Hu et al. \cite{hu2023point2partvolume} developed a method (Point2PartVolume) based on DL to predict both the whole body and parts' segment volumes of dressed subjects, using a single depth image. It is based on a two-step training strategy: the first to complete the partial body point cloud, predict the undressed body shape, and segment the body into six segments (head, torso, arms, and legs); the second to estimate the whole and partial volumes. They trained their method with synthetic data and tested it both on synthetic and on real data using BUFF \cite{zhang2017detailed} and PDT13 datasets \cite{helten2013personalization}. 

However, evaluating the human body volume with a single RGB-D camera is an ill-posed problem and cannot reach enough accuracyfor healthcare applications where precision is essential for accurate diagnosis and effective patient monitoring.
In addition, the subjects should wear tight clothes or just underwear to be able to estimate volumes with high precision. %, and this is not an issue in a medical office.
Indeed, Garcia Flores et al. \cite{garcia2024development}, aiming at the estimation of body volume and fat mass, reconstructed a 3D human body model using two Kinect cameras, one placed in front and one on the back of subjects wearing just underwear. The body volume estimations were compared to those obtained with the air displacement plethysmography. However, even if they obtained acceptable full-body volume estimations, the precision is limited by the Kinect resolution, and they did not estimated the segment volumes. %They showed that their body volume measurements are slightly underestimated maybe to the lack of lateral information. 
To the best of our knowledge, no work at the SoA employs a limited, but sufficient number of RGB-D cameras to obtain high estimation accuracy of both full-body and segment volumes.
In addition, even if part volumes are considered, the limbs are considered as single entities. However, since the ratio between proximal and distal parts change with subject health, many clinical applications, such as heart failure, lymphedema, and diabetes diagnosis, would benefit the knowledge of the distal parts volumes \cite{wilson2013ratio, hattori2021upper, bluher2019new}.
Moreover, the accuracy of the 3D reconstructed models is limited by the low resolution of the Kinect camera (Section \ref{subsec:3DScanTech}) that is the only depth camera tested.
Therefore, in this paper we present a method to accurately estimate the full-body and segment volumes, including proximal and distal parts of the limbs, employing front and back RGB-D camera views.
%Therefore, in this paper we present a system composed by two LiDAR RGBD cameras placed in front and on the back of the subject to accurately estimate the full-body volumes and segment volumes, including proximal and distal parts of the limbs.

%\hcor{Solo da RGB}
%Leinen et al. \cite{leinen2021volnet} developed VolNet, a deep neural network architecture to estimate human body segment volumes from a single RGB image. It is composed of different stages: first, the 2D pose is predicted, then the human body image is segmented in 14 human body parts, so as to estimate the 3D pose, and finally predict the volumes.

% Articoli che fanno stima volumi
% "Assessment of Body Volume Using Three-Dimensional Photonic Scanning"
% "Automated body volume acquisitions from 3D structured-light scanning"
% "Validation of a 3-dimensional photonic scanner for the measurement of body volumes, dimensions, and percentage body fat"
% "Human Body Volume Estimation in a Clinical Environment"
% "Volumeter: 3D human body parameters measurement with a single kinect"
% "Point2PartVolume: Human Body Volume Estimation From a Single Depth Image"
% "Using the microsoft kinect for patient size estimation and radiation dose normalization: Proof of concept and initial validation"
% "Development and Validation of a Method of Body Volume and Fat Mass Estimation Using Three-Dimensional Image Processing with a Mexican Sample"

% da RGB
    % "VolNet: Estimating Human Body Part Volumes from a Single RGB Image"

\section{BSV estimation pipeline} \label{sec:BSV}

%The BSV estimation pipeline presented here is specifically developed for healthcare applications where obesity based on individual body shape needs to be assessed, such as in the early diagnosis of heart failure. Its strengths are the ease of installation, low cost, and high precision.
The BSV estimation pipeline presented is specifically developed for applications where a 3D human body model with high accuracy is needed.
The system is composed of two LiDAR RGB-D cameras facing each other, among which the subject must stand still, turned towards one camera. %must stand in A-pose, turned towards one camera, with both palms forward-facing.
After the acquisition of both RGB and depth of front and back views of the subject, RGB pictures are employed for landmark detection and body part segmentations.
Then, depth images are aligned with RGB and converted to the corresponding texturized PCs, which are cleaned if needed and merged to obtain a unique PC for each subject.
Due to the lack of overlapping between the front and back acquisitions, the resulting PC is characterized by lateral gaps. Thus, a non-rigid regularized registration algorithm is employed to register the PC to a mesh template to compute the 3D subject mesh from which the whole body volume can be estimated. 
In addition, the volumes of the watertight segment meshes are obtained after the mesh segmentation and the closure of the different part meshes.
%Figure \ref{fig:Setup} and \ref{fig:SWpipeline} show an overview of the setup of the BSV estimation system and the software pipeline, respectively.
\begin{figure}[!h]
\centering
\includegraphics[width=3in]{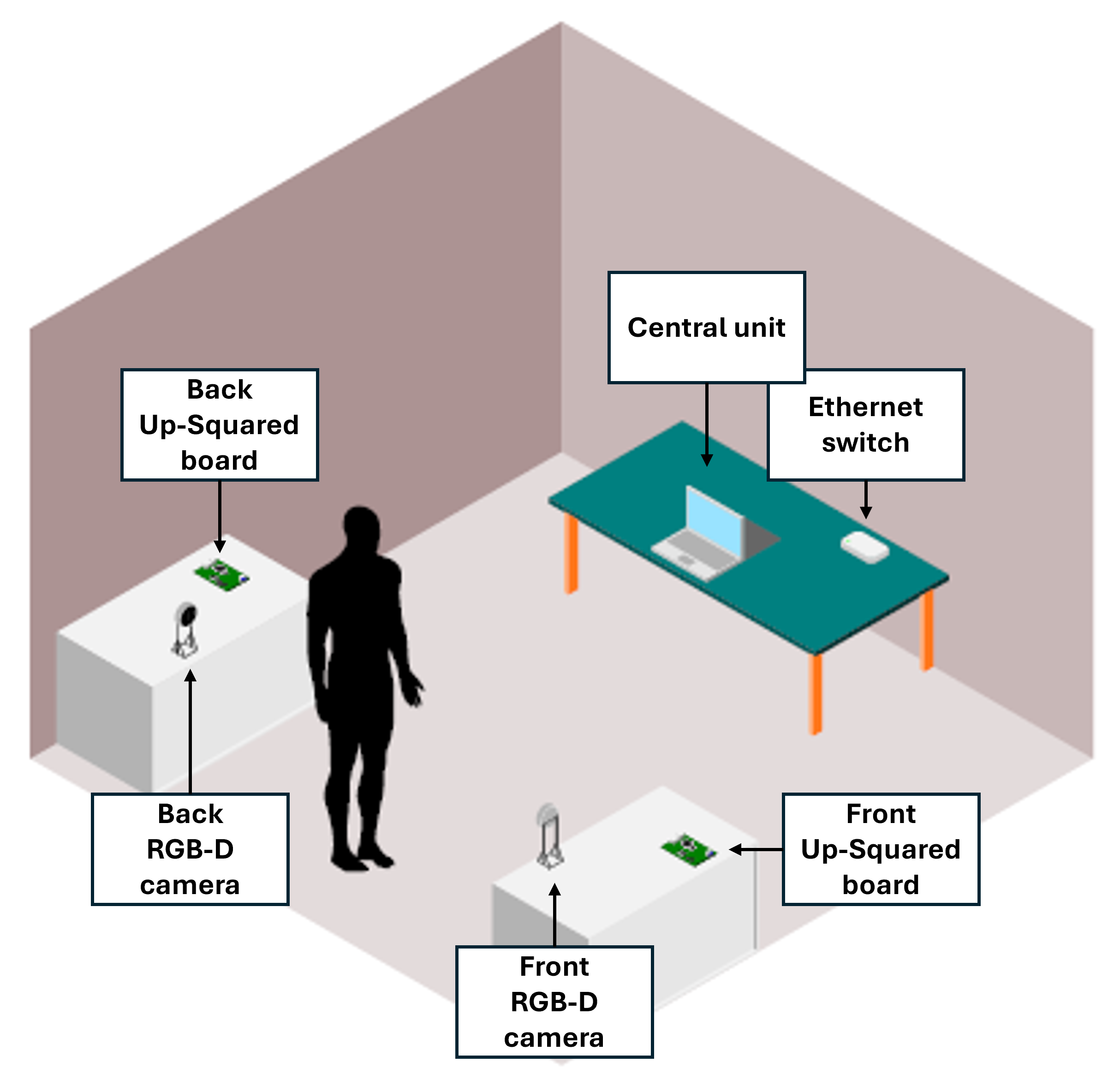}
\caption{BSV estimation system setup}
\label{fig:Setup}
\end{figure}
%\hcor{The overview of the method pipeline is presented in Figure \ref{}}

%Quindi vengono estratte le due point clouds front e retro dalle quali è estratta l'immagine RGB che viene usata per la segmentazione. Quindi vengono create le due point clouds texturizzate che poi vengono mergiate per ottenere la point cloud del soggetto alla quale si eliminano gli elementi distali (testa, mani e piedi).A questo punto si fa la ricostruzioe del modello 3D con l'algoritmo di registrazione non rigida basata su ARAP usando un template unico per ottenere la mesh del soggetto a partire dalla point cloud con gap laterali.Quindi si fa la segmentazione della mesh, si creano le watertight meshes e si calcolano i volumi dell'intero corpo e dei vari segmenti.

%The pipeline presented estimates the whole-body and part volumes using both RGB and depth front and back views of the subject. 
\subsection{System setup}

\begin{figure}[!h]
\centering
\includegraphics[width=3.5in]{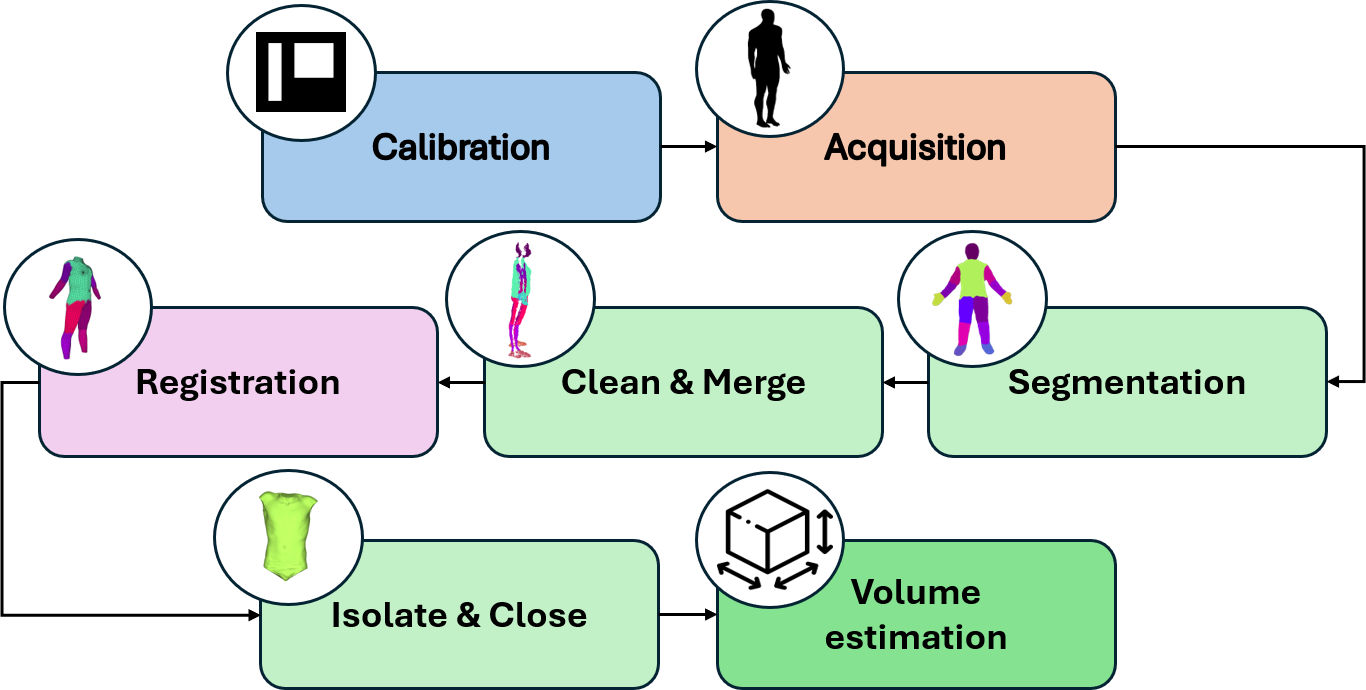}
\caption{BSV estimation software pipeline}
\label{fig:SWpipeline}
\end{figure}

As presented in Section \ref{subsec:3DScanTech}, LiDAR technology provides precise and detailed depth maps. Thus, we employed two Intel\textcopyright RealSense\textsuperscript{TM} LiDAR RGB-D camera L515 with depth resolution of $1024$ x $768$ pixels and accuracy \textless$5$ mm at $1$ m, and RGB resolutions of $1920$x$1080$ pixels. % which include also an RGB imager with a $1920$x$1080$ pixels image format. 
The L515 cameras are managed by two dedicated UP Squared boards, which are connected via SSH through an Ethernet switch to a central unit for user interface and data processing placed directly on the desk (Figure \ref{fig:Setup}).
The software pipeline (Figure \ref{fig:SWpipeline}) is implemented in Python 3 and described in the following Sections.

\subsection{Calibration and Acquisition} \label{sec:CalAcq}

%To establish a common reference frame when merging the captured point clouds, we performed a preliminary calibration by placing an ArUco marker at the center of the captured scene between the two aligned cameras.
Systems composed of multiple depth cameras need proper calibration to achieve accurate results and merge the captured PCs. Calibration involves aligning the cameras by transforming their local coordinate systems into a single and shared reference framework before scanning. 
This process requires determining both intrinsic and extrinsic parameters. 
%\hcor{The former describes the internal characteristics of each camera. The latter defines the spatial relationship and transformation from the camera’s local coordinate frames to the global reference system.}
Thus, we performed a preliminary calibration placing an ArUco marker (”Original ArUco” dictionary, ID 336, 16×16 cm) in the center of the captured scene, as a common reference frame, to compute the parameters when the two cameras are aligned.
To this end, we developed an interactive calibration procedure that displays, in real-time, the captured ArUco marker, the computed rotation around the longitudinal axis, and the translation along the transversal axis that each camera must undergo to be aligned with it, and notifies when the camera is aligned. 
At the end of the calibration, the cameras can be secured and the system is ready for the acquisition procedure, which requires the patient to stand still for 5 seconds in A-pose, turned towards one camera, with both palms facing forward (Figure \ref{fig:Setup}).

%ArUco is an open-source library for detecting square fiducial markers in images. If the camera is calibrated, you can estimate the pose of the camera with respect to the markers. The library is written in C++, but a python binder is available, and requires OpenCv. 

%The L515 camera features a factory calibration, and the intrinsic parameters consisting in a camera matrix and 5 parameters Brown - Conrady distortion model are saved in the device and can be accessed. Once retrieved it is possible to get the transformation from the marker frame to the camera frame.

%\subsection{Software architecure}

\subsection{Landmark Detection and Segmentation} \label{sec:Seg}

To estimate the volumes of each body part, the body must be segmented.
We considered $14$ segments, featuring $2$ central segments: head and torso, and $12$ distal segments: left and right arms, forearms, hands, thighs, shins, and feet.
% (Figure \ref{fig:BodySegmentswithextr}).
However, since we want to estimate the volume ratio between body segments with variations in adiposity, the head, hands, and feet segments are discarded immediately before the registration phase. % (Figure \ref{fig:BodySegmentsNoextr}).
They do not add significant information that could justify the growth of computational load due to the need for a different registration methodology.
% They do not add significant information compared to the difficulties they pose in reconstruction. 
%\hcor{Since, unlike the torso region and limbs, which can be processed together through a single algorithm featuring the same parameters, they would require different registration methodologies due to their complexity, high shape variability, and lower point density when capturing such small areas.}

To further limit the computational cost, we used RGB data for feature extraction, rather than working with more complex reconstructed PCs. We employed MediaPipe (MP) \cite{lugaresi2019mediapipe} for landmark identification and BodyPix (BP) \cite{bodypix2019} for body segmentation.
%With the same purpose of minimizing the computational resources needed during processing, we used RGB data for feature extraction.
%This kind of design allows for the deployment of fast, reliable, and lightweight solutions such as Deep Neural Networks (DNN), rather than working with more complex reconstructed PCs. 
%In particular, we employed MediaPipe \cite{lugaresi2019mediapipe} for landmark identification and BodyPix \cite{bodypix2019} for body segmentation.

Landmark detection allows both to check if all features of interest are captured during the acquisition phase and to check and eventually correct the segmentation output.
%MediaPipe is a framework for building Machine Learning (ML) perception pipelines to perform inference on multimodal sensor data through a graph-modular approach.
Among the different MP solutions proposed, we deployed the MP Holistic (MPH) pipeline to perform body landmark detection. MPH combines three distinct models for human pose, face, and hand landmarks, which provide $33$, $468$, and $21$ landmarks respectively, for a total of $543$ landmarks.
%%Because of their specialized designs, the input requirements for one component are not ideal for the others. Therefore, MPH functions as a multi-stage pipeline that processes each region using a resolution suited to its specific requirements.
%MPH combines three distinct models for human pose, face, and hand landmarks, each tailored to its specific domain.
%MPH pose, face, and hand landmark models provide $33$, $468$, and $21$ landmarks respectively, for a total of $543$ landmarks.
However, since we are not interested in the head segment, MPH face landmarks are not considered. We also discarded the right and left thumbs and wrists of the pose and hand modules, respectively, to reduce the redundancy and number of landmarks, obtaining a total of $71$ body landmarks.
In addition, MPH allows the tuning of a landmark detection confidence parameter ranging from $0$ to $1$. We set the pipeline to work with a $0.75$ confidence level as it proofs sufficiently robust and accurate.

\begin{figure}[t]
     \centering
     \begin{subfigure}[h]{0.49\columnwidth}
        \centering
         \includegraphics[width=1.5in]{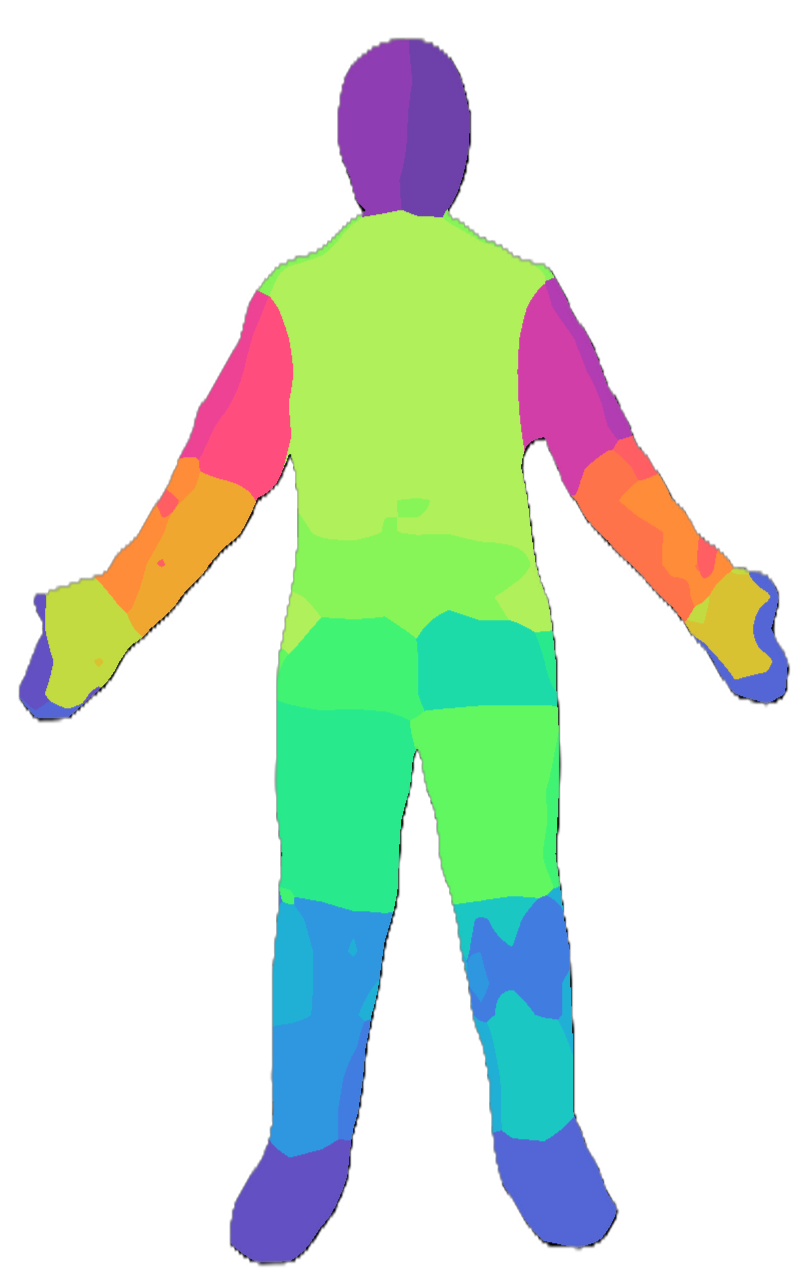}
         \caption{Segment classification output of BodyPix.}
         \label{fig:BPerrors}
     \end{subfigure}
     \hfill
     \begin{subfigure}[h]{0.49\columnwidth}
         \includegraphics[width=0.88\columnwidth]{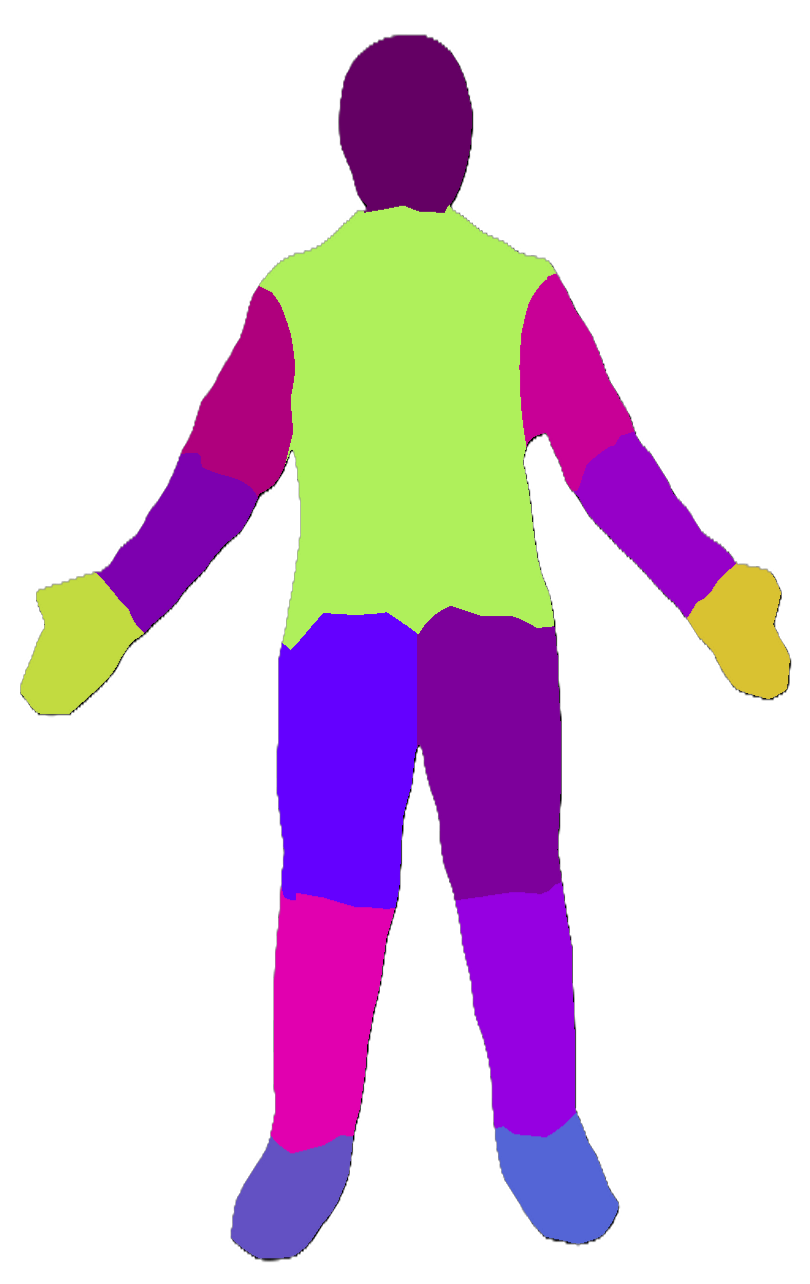}
         \caption{Segment classification re-labelled.}
         \label{fig:BPrelabeled}
     \end{subfigure}
    \caption{Segment classification output of BodyPix before (a) and after (b) the correction.}
    \label{fig:BodyPix}
\end{figure}

%BP is an open-source ML model based on ResNet-50 architecture. 
BP is specifically trained to first segment the image into pixels that belong or not to a person (BP mask) and to further classify the pixels representing the person into $24$ body parts.
However, the segmentation output of BP is not robust enough for our application. Even if it retains valuable segmentation information, it always exhibits the same predictable errors: parts of the left and right analogous segments and hands and feet extremities are often mixed (Figure \ref{fig:BPerrors}). %(e.g., areas of the right lower leg are mistakenly labelled as left lower leg, and vice versa) (Figure \ref{fig:BPerrors}).
Thus, we corrected the BP output by integrating the information previously obtained in the landmark detection phase. 
First, we computed a new segmentation by grouping the original BP $24$ segments into the $14$ segments. 
Then, the medial and transverse lines of the body are estimated and the segments are then re-labelled accordingly to the plane portion they occupy with respect to these references (Figure \ref{fig:BPrelabeled}).
%The first, which corresponds to the intersection of the sagittal plane, is evaluated by computing the line parallel to the y-axis in the image frame crossing the medial point of the segment connecting the right and left hip landmarks.  
%The second, which corresponds to the intersection of the transverse planes, is evaluated by computing the line parallel to the x-axis in the image frame crossing the medial point of the segment connecting the right and left knee landmarks.
%The segments are then re-labelled accordingly to the plane portion they occupy with respect to these references (Figure \ref{fig:BPrelabeled}).
%\hcor{Alla fine ho preferito per l'utilizzo della segmentazione solo del front per ridurre i problemi di allineamento. Si dice o si lascia così?}
At the end of the segmentation process, we texturized the front and back 3D PCs with the computed segmentation RGB images. % computed \hcor{using the deprojection features included in the .bag file saved during the acquisition and the pyrealsense2 library provided by Intel\textcopyright RealSense\textsuperscript{TM}}.

%noi grazie all'uso di RGBD camera possiamo sfruttare l'RGB

\subsection{Point Cloud Cleaning and Merging} \label{sec:CleanMerge}

The generated front and back PCs needed to be cleaned.
First, to remove other objects present around the human body, we isolated the body PC by removing the points falling outside the silhouette provided by the BP mask. %Thus, we isolated the body point cloud by removing the points falling outside the silhouette provided by the segmentation mask.
Then, since the remaining PC is usually affected by outliers, mainly due to errors in depth detection, we performed a statistical outlier removal, with $600$ neighbors and a standard deviation ratio of $0.05$, to prune the PC from the points deviating significantly from the whole body position in space.
%\hcor{Open3D library provides a ”statistical outlier removal” function, that removes points that are further away from their neighbors compared to the average for the point cloud. It takes two input parameters: the number of neighbours taken into account in order to calculate the average distance for every given point, and a parameter based on the standard deviation of the average distances across the point cloud. The lower this number, the more aggressive the filter will be. Here we perform outlier removal with $600$ neighbors and a standard deviation ratio of $0.05$.}

After the cleaning, the front and back PCs are merged to form a unified PC by applying the rigid transformations previously computed in the preliminary calibration phase (Section \ref{sec:CalAcq}).% to express the front and back captured points in the marker frame. 
The result is a full-body segmented PC characterized by lateral gaps between the front and back PCs (Figure \ref{fig:UnifiedPC}). 
\begin{comment}
\begin{figure}[!t]
\centering
\includegraphics[width=1in]{Pics/side_seg_pcd.png}
\caption{\hcor{Inserire caption}}
\label{fig:UnifiedPC}
\end{figure}
\end{comment}
Finally, the head, hands, and feet are excluded from the PC of the entire body.

%each segment's point clouds are isolated and saved separately, and then recombined into a unified point cloud featuring torso, arms, forearms, and upper and lower legs, discarding head, hands, and feet segments (\hcor{Figure \ref{}}) 

\subsection{3D non-rigid registration algorithm} \label{seq:Registration}

As previously presented, we employed a non-rigid registration technique to warp a template on the 3D PC obtained in the previous steps.
%The template-based body model registration method ensures geometric coherence in the gap areas. 
%Thus, we created a generic human template mesh using MakeHuman\textsuperscript{TM}, which is an open-source software dedicated to the creation of humanoid prototypes for 3D computer graphics. It allows the selection of different grades of body features such as sex, age, adiposity, and height, and, for the sake of simplicity, we selected a single template with average characteristics to fit both sexes and different body types with a single mesh. The template mesh is then manipulated in Blender, an open-source software for modelling, rigging, and animating 3D models, to watertight the mesh, put it in a similar pose to the one kept during the acquisition, and remove the head, hands, and feet. The mesh template must be scaled and aligned to the 3D point cloud. Therefore, we translated both of them to the center of the scene and computed the scaling factor as the ratio between the lengths of the bounding boxes' diagonals. Then, we aligned the bounding box of the scaled mesh to that of the point cloud.
We created a generic human template mesh, with average characteristics to fit both sexes and different body types with a single mesh, using MakeHuman\textsuperscript{TM}.
Then, we employed Blender to put it in a similar pose to the one kept during the acquisition, remove the head, hands, and feet, and watertight the mesh.
%We created a generic human template mesh using MakeHuman\textsuperscript{TM}, which is an open-source software that allows the selection of different grades of body features such as sex, age, adiposity, and height of humanoid prototypes. For the sake of simplicity, we selected a single template with average characteristics to fit both sexes and different body types with a single mesh. Then, we employed Blender, an open-source software for modelling 3D models, to put it in a similar pose to the one kept during the acquisition, remove the head, hands, and feet, and watertight the mesh.
Next, we scaled and aligned the mesh template to the 3D PC, translating both of them to the center of the scene, computing the scaling factor as the ratio between the lengths of the bounding boxes' diagonals, and finally aligning the two bounding boxes.

After the alignment phase, we performed the non-rigid deformation of the template mesh in order to fit the 3D PC. As presented in Section \ref{seq:Regmethods}, these methods compute the deformation by optimizing a target energy function. We employed the ARAP deformation algorithm \cite{sorkine2007rigid} as the fitting term, and with the aim of preserving the original features of the template and providing a meaningful reconstruction of the lateral gaps in the PC, we introduced a regularization term inspired by the one presented in \cite{kwok2014volumetric}.
ARAP is based on the concept of local rigidity and states that given a triangular mesh \textit{S} with \textit{n} vertices \textbf{p} and \textit{m} triangles, it is possible to find a new mesh \textit{S'}, with vertices \textbf{p'}, that is locally deformed as rigid as possible so that it is not stretched, flattened, or sheared. Thus, they look for the global deformation that minimizes the divergence of the cells deformation $\textbf{R}_{i}$ from being rigid. The ARAP energy guarantees the preservation of rigidity in a least-squares sense, and it is expressed as follows:
\begin{equation} \label{eq:ARAP}
E_{ARAP} =  \sum_{i=1}^{n}w_{i}\sum_{j\in\cal{N}(i)} w_{ij}\parallel(\textbf{p}_{i}'-\textbf{p}_{j}')-\textbf{R}_{i}(\textbf{p}_{i}-\textbf{p}_{j})\parallel^2
\end{equation}
where $\textbf{p}_{i}$ and $\textbf{p}_{j}$ are the vertices of an edge of the undeformed mesh, which are deformed into the vertices $\textbf{p}_{i}'$ and $\textbf{p}_{j}'$, $w_{i}$ and $w_{ij}$ are the per-cell and per-edge weights, respectively, and $\textbf{R}_{i}$ is the rotation matrix to transform the cells composing the surface. 
The per-edge weights $w_{ij}$ need to compensate for the influence of the meshing bias. Thus, to reduce asymmetric deformations, they employed the cotangent weighting factor defined as follows:
\begin{equation} \label{eq:wij}
w_{ij} =  \frac{1}{2}(cot\alpha_{ij}+cot\beta_{ij})
\end{equation}
where $\alpha_{ij}$ and $\beta_{ij}$ are the opposite angles of the mesh edge.

Minimizing $E_{ARAP}$ with respect to $\textbf{R}_{i}$ and $\textbf{p}'$, alternatively, a local energy minimum is reached. To derive the optimal rotation $\textbf{R}_{i}$ keeping $\textbf{p}'$ fixed, they considerd the edges $\textbf{e}_{ij}=\textbf{p}_{i}-\textbf{p}_{j}$ and $\textbf{e}_{ij}'=\textbf{p}_{i}'-\textbf{p}_{j}'$ and derived $\textbf{R}_{i}$ as the Singular Value Decomposition (SVD) of the covariance matrix $\textbf{S}_{i}=\textbf{U}_i\Sigma_{i}\textbf{V}_{i}^T$:
\begin{equation} \label{eq:Ri}
R_{i} =  V_{i}U_{i}^T
\end{equation}
finding the smallest singular value such as $det(\textbf{R}_{i})$ \textgreater $0$.
Once the value for $\textbf{R}_{i}$ is established, the position  $\textbf{p}'$ must be computed from a given $\textbf{R}_{i}$, they computed the gradient of the energy $\textbf{E}_{ARAP}$ with respect to $\textbf{p}'$ and derived the following linear system of equations:
\begin{equation} \label{eq:p}
\sum_{j\in\cal{N}(i)}w_{ij}(\textbf{p}_{i}'-\textbf{p}_{j}') = \sum_{j\in\cal{N}(i)} \frac{1}{2}(\textbf{R}_{i}+\textbf{R}_{j})(\textbf{p}_{i}-\textbf{p}_{j})
\end{equation}
where the left-hand side is the discrete Laplace-Beltrami operator applied to $\textbf{p}'$, which is equal to \textbf{b}, an \textit{n}-vector whose each rows contains the right-hand expression of \ref{eq:p}:
\begin{equation} \label{eq:L}
Lp'=b
\end{equation}

\begin{figure}[h]
\centering
\includegraphics[width=3in]{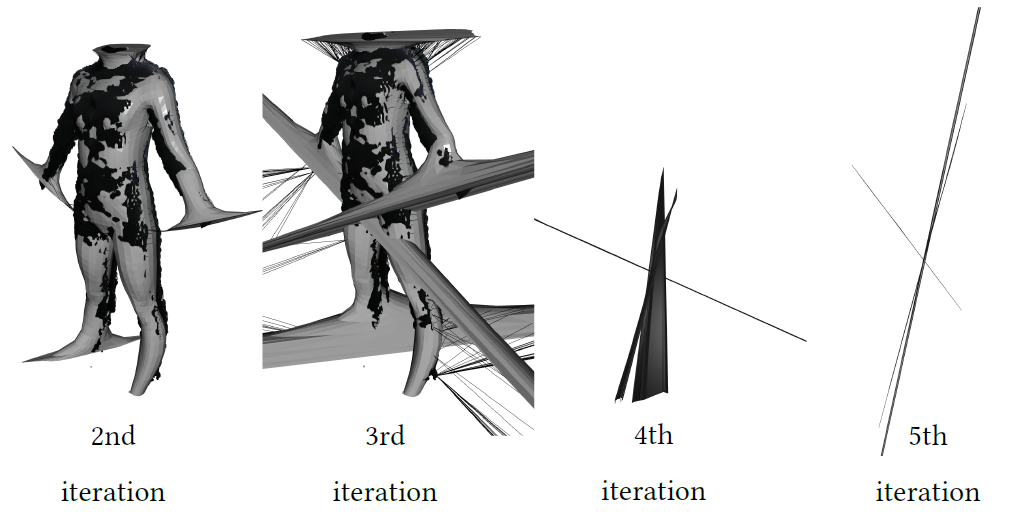}
\caption{Fitting results when registering with pure ARAP.}
\label{fig:collapse}
\end{figure}

However, without including a regularization term, perfect correspondences are found, resulting in undesirable mesh properties, such as rapidly changing local geometry. Figure \ref{fig:collapse} shows how the lack of regularization action causes incongruous deformation, leading to numerical instability after only $5$ iterations and ultimately failing reconstruction. ARAP is fast failing because there is no rejection of point clustering, and the absence of counterweighting of the deformation effect causes each vertex to undergo the same deformation.
Thus, to preserve the original spatial features of the mesh template, such as angles, areas, volumes, and edge lengths, we introduce the regularization term defined as follows:
\begin{equation} \label{eq:Regterm}
E_{regularization} =  \alpha A \sum_{j\in\cal{N}(i)}\parallel U_i\Sigma_iV_i^T -  U_iV_i^T \parallel_F^2
\end{equation}
where $\parallel . \parallel_F$ is the Frobenius norm, $U_i\Sigma_iV_i^T$ is the SVD of $\textbf{R}_{i}$, $\alpha$ is a weighting parameter regulating the trade off between fitting and preserving the original geometric features of the template mesh, and $A$ is the total area of the mesh. 
Considering the total area of the mesh instead of the area of each triangle leads to the reset of the regularization energy when the triangles collapse. 
Thus, the final form of the energy to be minimized is:
\begin{equation} \label{eq:Ecomp}
E(S') =  \sum_{i} w_{i} (E_{fitting} + E_{regularization})
\end{equation}

\begin{align*}\label{eq:E}
E(S') =  \sum_{i=1}^{n}w_{i}(\sum_{j\in\cal{N}(i)} w_{ij}\parallel(\textbf{p}_{i}'-\textbf{p}_{j}')-\textbf{R}_{i}(\textbf{p}_{i}-\textbf{p}_{j})\parallel^2 + \\
\alpha A \sum_{j\in\cal{N}(i)}\parallel U_i\Sigma_iV_i^T -  U_iV_i^T \parallel_F^2)
\end{align*}
where we set the per-cell weight to $10^-2$ and $\alpha$ to $10^6$.

In order to compute the optimal transformation, every mesh vertex must be assigned to a target position, i.e., the corresponding point on the PC.
These correspondences are generally found by Nearest Neighbor (NN) techniques. This can be approached in two opposite directions of the NN search: mesh-to-point (m2p) \cite{li2009robust}, or point-to-mesh (p2m) \cite{zell2013elastiface}. If the PC is uniformly distributed, the m2p approach gives a good fit. However, when the PC presents gap areas, m2p can cause undesired deformations when fitting the gaps' proximity. In this case, the p2m approach can overcome this problem, but it does not align the mesh to the PC as well as the m2p technique.
For these reasons, we adopted the following strategy: we implemented the first $5$ ARAP iterations with the m2p approach, then $5$ iterations with the p2m approach, and finally $10$ iterations with the m2p approach.

\subsection{Mesh Segmentation and Volume Estimation}

\begin{figure}[t]
     \centering
     \begin{subfigure}[h]{0.49\columnwidth}
        \centering
         \includegraphics[width=0.7in]{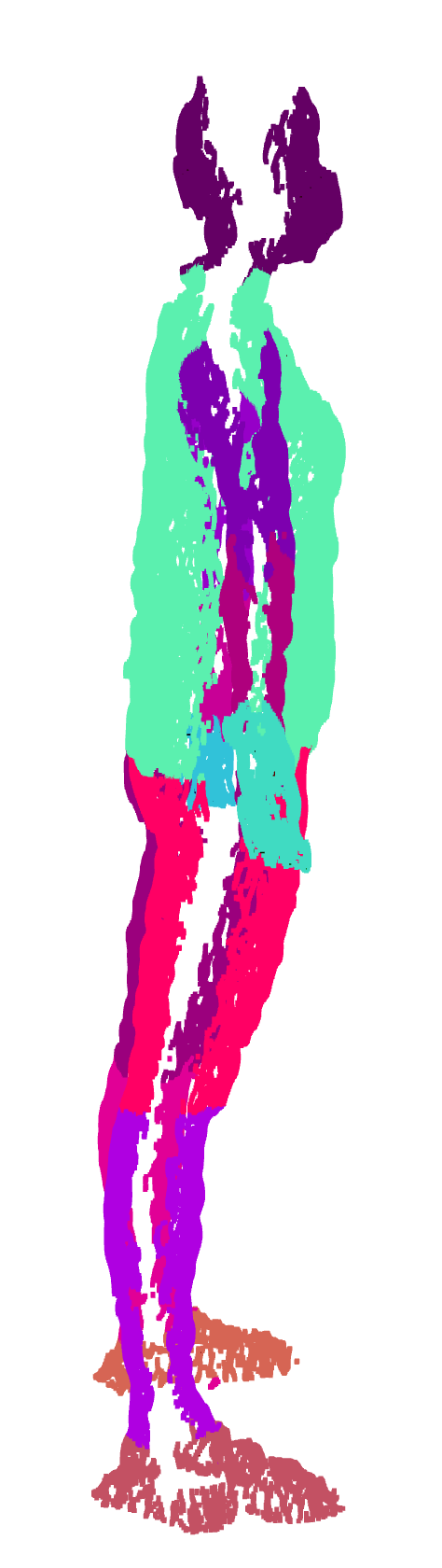}
         \caption{Segmented point cloud with lateral gaps.}
         \label{fig:UnifiedPC}
     \end{subfigure}
     \hfill
     \begin{subfigure}[h]{0.49\columnwidth}
         \includegraphics[width=0.67\columnwidth]{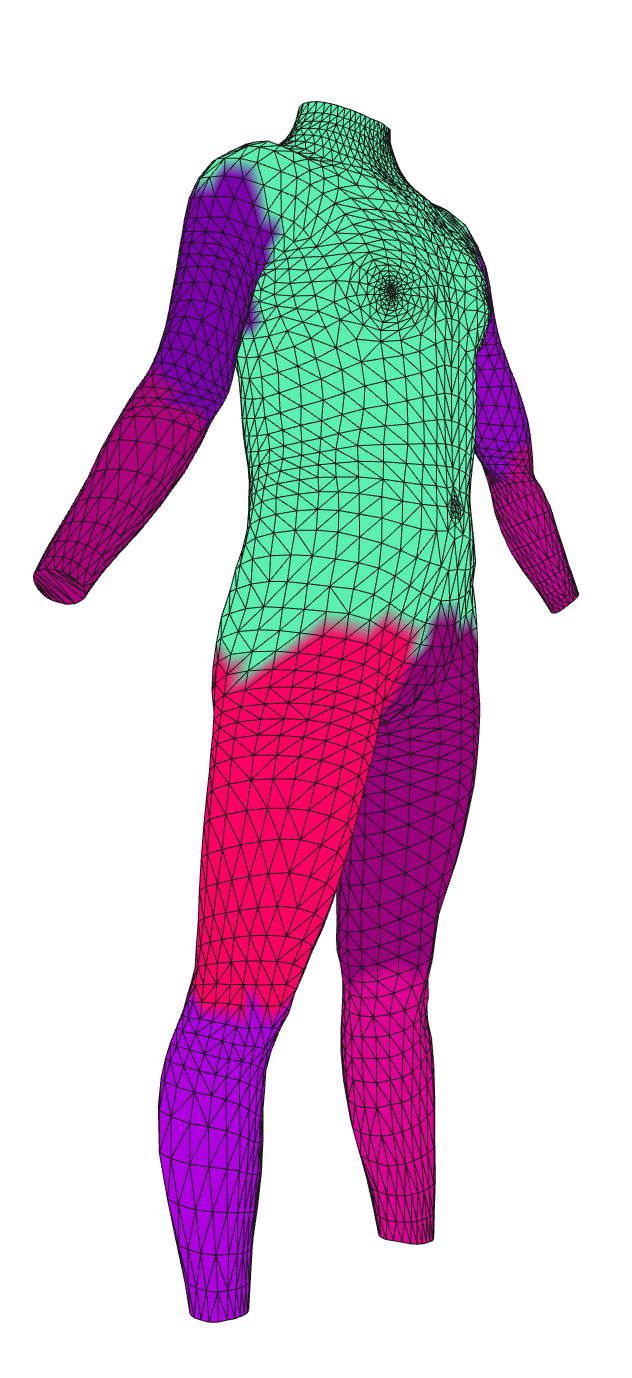}
         \caption{Fitted segmented mesh.}
         \label{fig:SegMesh}
     \end{subfigure}
    \caption{Full-body point cloud before removing extremities (a) and segmented mesh after registration algorithm (b).}
    \label{fig:pcdmesh}
\end{figure}

We obtained the segmented mesh (Figure \ref{fig:SegMesh}) labelling each vertex of the resulting fitted mesh as its NN on the PC, and we isolated each segment mesh, which are then watertight, to finally estimate both the full-body and segment volumes.% in MeshLab, an open source software for processing and editing 3D triangular meshes.
\begin{comment}
\begin{figure}[h]
\centering
\includegraphics[width=1.3in]{Pics/3_4_seg_mesh.png}
\caption{\hcor{Inserire caption}}
\label{fig:SegMesh}
\end{figure}
\end{comment}

\section{Validation} \label{sec:Validation}

BSV is assessed with five consecutive evaluation steps.
We evaluated the L515 camera errors during the calibration procedure. We assessed the goodness of the 3D non-rigid registration algorithm, estimating the volume of two known objects with different form factors, and then evaluating the full-body volume and the $9$ body parts volumes employing the FAUST dataset \cite{bogo2014faust}, a dataset especially developed for the evaluation of 3D mesh registration algorithms. 
Then, we also compare our results to those obtained by Hu et al. in \cite{hu2023point2partvolume}, who predicted $6$ partial volumes using a single depth image (Section \ref{seq:VolumeEstimate}). 
Finally, we applied BSV to real acquisition data.

\subsection{Calibration evaluation} \label{seq:calibeval}

% Errori di orientazione e traslazione stimati

To estimate the L515 camera errors, we set up the BSV system in a controlled manner. We placed the front and back RGB-D cameras in the same positions and orientations using reference points placed on the ground and placing them at the same height. 
In particular, the front and back cameras were $400.6 cm$ apart from each other, at a height of $100.3 cm$ and $99.9 cm$, and at $+0.34 cm$ and $-0.41 cm$ with respect to the transverse axis, respectively. This opposite displacement of the cameras is due to their rectangular aspect ratio and the consequent need to turn them $90^{\circ}$ to capture the full shape of the human body, resulting in the RGB imager not being aligned with the longitudinal axis.
In addition, we placed the ArUco marker in the center of the line of sight of the two cameras at the same distance from them ($200.1 cm$ and $200.5 cm$ from the front and back cameras, respectively).
This setup allows evaluating the camera errors in estimating the extrinsic parameters, that is, the translations and rotations from the cameras' local coordinate frames to the marker global reference system. With this aim, we made five consecutive acquisitions to also evaluate the repeatability of the calibration procedure.

\subsection{3D non-rigid registration algorithm evaluation} \label{seq:voleval}

\subsubsection{Known objects volume evaluation} \label{seq:boxeval}
% Ricostruzione box 

To assess the goodness of the 3D non-rigid registration algorithm, we first estimated the volume of two boxes with different form factors. The height, width and depth of one box (Box 1) were $55.8 cm$, $52 cm$, and $58.9 cm$, respectively, resulting in a volume of $0.171m^3$. For the other box (Box 2) were $103.8 cm$, $20.8 cm$, and $20.4 cm$, respectively, resulting in a volume of $0.044m^3$.
We placed the two boxes at $45^{\circ}$ with respect to the line of sight of the front and back cameras to allow them to capture all three dimensions and we made three consecutive acquisitions.

\subsubsection{3D Human Body Model reconstruction evaluation} \label{seq:humaneval}

\begin{figure}[t]
     \centering
     \begin{subfigure}[h]{0.49\columnwidth}
        \centering
         \includegraphics[width=0.83in]{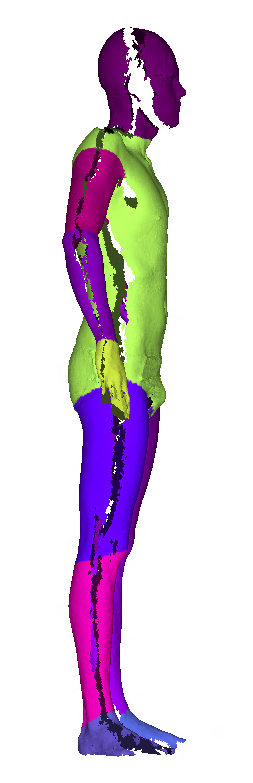}
         \caption{Segmented PC with lateral gaps.}         \label{fig:LateralSegmentedPointCloudFAUST}
     \end{subfigure}
     \hfill
     \begin{subfigure}[h]{0.49\columnwidth}
         \includegraphics[width=0.7\columnwidth]{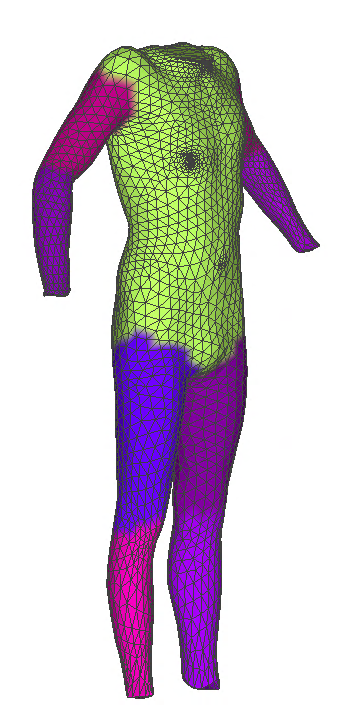}
         \caption{Fitted segmented mesh.}
         \label{fig:FittedFAUST}
     \end{subfigure}
    \caption{Full-body PC before removing extremities (a) and segmented mesh after registration algorithm (b) of subject $1$ of FAUST dataset.}
    \label{fig:FAUST}
\end{figure}

After having evaluated the 3D registration algorithm with objects with known volumes, estimations of the full-body and part volumes must be compared to a gold standard.
Nowadays, the most affordable and appropriate way to validate a new 3D human body model reconstruction algorithm is the use of a publicly available dataset \cite{deng2022survey}.
In particular, the FAUST datasets \cite{bogo2014faust} is specifically designed as a 3D mesh registration algorithm benchmark. 
Thus, we used it to compare full-body and body parts volume estimations obtained on the original FAUST meshes and those obtained with BSV.
FAUST contains high-resolution (approximately $180k$ vertices and more than $300k$ triangles) human scans of $10$ subjects in $30$ different postures with ground-truth correspondances. The scans were acquired by a full-body 3D stereo capture system (3dMD, Atlanta, GA) composed of $22$ 3D multi-stereo cameras and they achieved accurate template registration ($2mm$), using a dense texture pattern painted on the bodies.
%Since the FAUST dataset is specifically designed as a 3D mesh registration algorithm benchmark, we used it to compare full-body and body parts volume estimations obtained on the original FAUST meshes and those obtained with the BSV estimation pipeline.
%%Others \cite{cao2024motion2vecsets} focused in the reconstruction of time-varying surfaces from sequential point clouds and use Dynamic FAUST dataset \cite{bogo2017dynamic} which includes ground-truth correspondances and is perfectly suited for applications in human motion analysis and tracking.
%%Thus, in this work, we used the FAUST dataset because it is specifically designed as a 3D mesh registration algorithm benchmark.
%%However, since FAUST dataset only contains full-body 3D scans and, as presented in Section \ref{sec:Seg}, we used RGB images to identify landmarks and segment body parts, we slightly modify the BSV pipeline previously presented (Section \ref{sec:BSV}). In particular, we replaced the calibration and acquisition steps with an RGB front and back images extraction from the point clouds of the subjects using the Open3D library (Figure \ref{fig:FrontBackFAUSTPCimages}) to apply the MediaPipe and BodyPix algorithm to detect landmarks and segment body parts. 
However, since the FAUST dataset only contains full-body 3D scans, we slightly modified the BSV pipeline previously presented (Section \ref{sec:BSV}). 
First, we replaced the calibration and acquisition steps with a phase in which we extracted front and back RGB images and PCs from the full-body scans to simulate the use of two RGB-D cameras.
%\hcor{forse nn necessario specificare le librerie usate!}
%The RGB images are obtained with \textit{Open3D} library and we manipulated the color aspect with \textit{matplotlib} library (Figure \ref{fig:FrontBackFAUSTPCimages}) to allow MediaPipe and BodyPix algorithm to detect landmarks and segment body parts.  
The color aspect of the RGB images is manipulated to allow MP and BP algorithms to detect landmarks and segment body parts.  
\begin{figure}[h]
\centering
\includegraphics[width=3in]{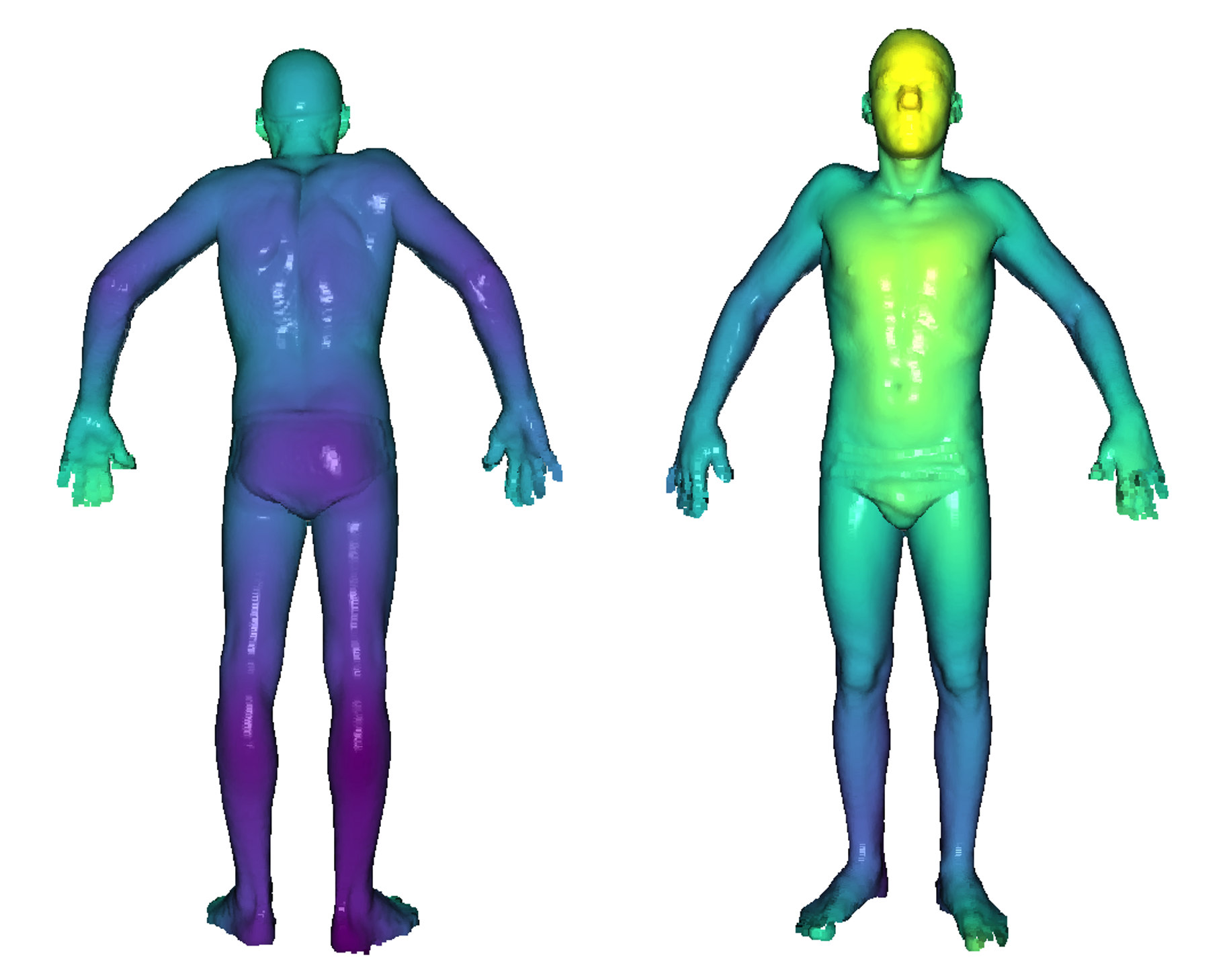}
\caption{Front and back RGB images of subject $1$ extracted from the FAUST dataset.}
\label{fig:FrontBackFAUSTPCimages}
\end{figure}
As showed in Figure \ref{fig:FrontBackFAUSTPCimages}, among the $30$ different poses present in the FAUST dataset, we selected the one most similar to the one that the subject should take during the acquisition with BSV. There are only two differences: the shoulders are kept up, and the palms are facing back, despite this, MP can detect all the landmarks.
Then, we projected the segmented output to the PC obtained by merging the front and back PCs. In this way, we got a segmented full-body point cloud with lateral gaps (Figure \ref{fig:LateralSegmentedPointCloudFAUST}) as the one attained with the original BSV (Section \ref{sec:CleanMerge}).
\begin{comment}
\begin{figure}[h]
\centering
\includegraphics[width=1in]{Pics/LateralSegmentedPointCloudFAUST.png}
\caption{\hcor{Inserire caption}}
\label{fig:LateralSegmentedPointCloudFAUST}
\end{figure}
\end{comment}
Then, we excluded the extremities and we employed the 3D non-rigid deformation algorithm to obtain the fitted mesh (Figure \ref{fig:FittedFAUST}) from which the segment meshes were isolated.
Finally, each full-body and segment mesh was automatically watertight, and the volume was estimated.

% CAMBIA Compared to other scanner types, they are the most accurate and reliable and are therefore typically used to obtain ground truth data, e.g., stationary scanners were used to create 3D body scanning datasets like CAESAR [3], SIZE-UK [8], Scan DB [64], and FAUST [24].
% "A Review of Body Measurement Using 3D Scanning"

%FAUST, NOMO3D, PDT13(?)

% Bogo et al. [4] introduced the FAUST dataset for comparing non-rigid registration methods.
% yan2020anthropometric

\subsection{Validation with real acquisitions}

Finally, we presented a qualitative validation of the BSV system making two real acquisitions with a male and a female subjects. 

\subsection{Evaluation metrics}

We evaluated the calibration errors calculating the difference between the real and the estimated values. In particular, for the translation errors, we had the real values, and for the orientation errors, we considered that the two cameras were perfectly facing each other thanks to the controlled setup.
For the evaluation of the 3D non-rigid registration algorithm, both in the case of known objects and FAUST body models, we used the Relative Volume Error (RVE) defined as:
\begin{equation} \label{eq:RVE}
RVE =  \vert(V_{est} - V_{GT})\vert/V_{GT}*100\%
\end{equation}
where $V_{est}$ is the estimated volume and $V_{GT}$ is the box volumes when considering the known objects, and the full-body or segment volumes of the original FAUST mesh when considering the human body models.
In the latter case, since the PCs extracted from the FAUST dataset can be considered error-free, we evaluated the RVE in four different conditions:
\begin{enumerate}
    \item camera calibration errors included (Cali);
    \item L515 depth error included (L515);
    \item both errors included (L5Ca);
    \item no errors included (NoEr).
\end{enumerate}
%We considered the situation in which we added camera calibration errors (Cali) (Section \ref{seq:calibeval}), the condition in which we added L515 depth error (L515), the condition when both errors are included (L5Ca), and the one when no errors are present (NoEr). We compared these four different volumes to the ground truth volume obtained from the original FAUST mesh.
In addition, we compared our results to those obtained by Hu et al. \cite{hu2023point2partvolume} because, to the best of our knowledge, Point2PartVolume is the most recent volume prediction method, which includes segment volumes estimation. However, as presented in Section \ref{seq:VolumeEstimate}, they segmented the body into only $6$ segments (head, torso, arms, and legs). Thus, we joined the mesh of the upper and lower parts of arms and legs to evaluate the limb volumes.
In this case, since Point2PartVolume is DL based, we compute the RVE accuracy as $100 - RVE$ to compare their accuracies results to ours.
Finally, when considering the BSV estimations of real acquisitions, we employed the Relative Mass Error (RME) computed as the RVE, but considering the real mass of the subjects and the estimated mass computed as the estimated full-body volume multiplied by the body density that is approximately $1000 Kg/m^3$.

\section{Results} \label{sec:Results}

\subsection{Calibration results}

Table \ref{tab:CalEr} presents the mean and standard deviation of the L515 camera translation ([cm]) and rotation ([deg]) errors in the camera reference system, where x is the vertical, y is the transversal, and z is the longitudinal axis.
\begin{table}[h]
    \caption{Mean and standard deviation (Std Dev) of the L515 camera translation [cm] and rotation errors [deg].}
    \centering
    \begin{tabular}{|l|l|l|l|}
        \hline
         \textbf{L515 camera}  & \textbf{Error} & \textbf{Mean} & \textbf{Std Dev} \\
         \hline
         Front & Translation & 0.22 0.53, -7.01 & 0.11, 0.03, 0.06 \\
         Front & Rotation    & 0.88, 0.68, 0.47 & 0.03, 0.01, 0.21 \\
         Back  & Translation & -0.17, -0.13, -5.84 & 0.15, 0.02, 0.34 \\         
         Back  & Rotation    & 0.24, 0.46, 1.05 & 0.28, 0.11, 0.10 \\
         \hline
    \end{tabular}
    \label{tab:CalEr}
\end{table}

%5.22
\subsection{3D models registration results}

\subsubsection{3D Box Model registration results}

\begin{figure}[h]
\centering
\includegraphics[width=0.99\linewidth]{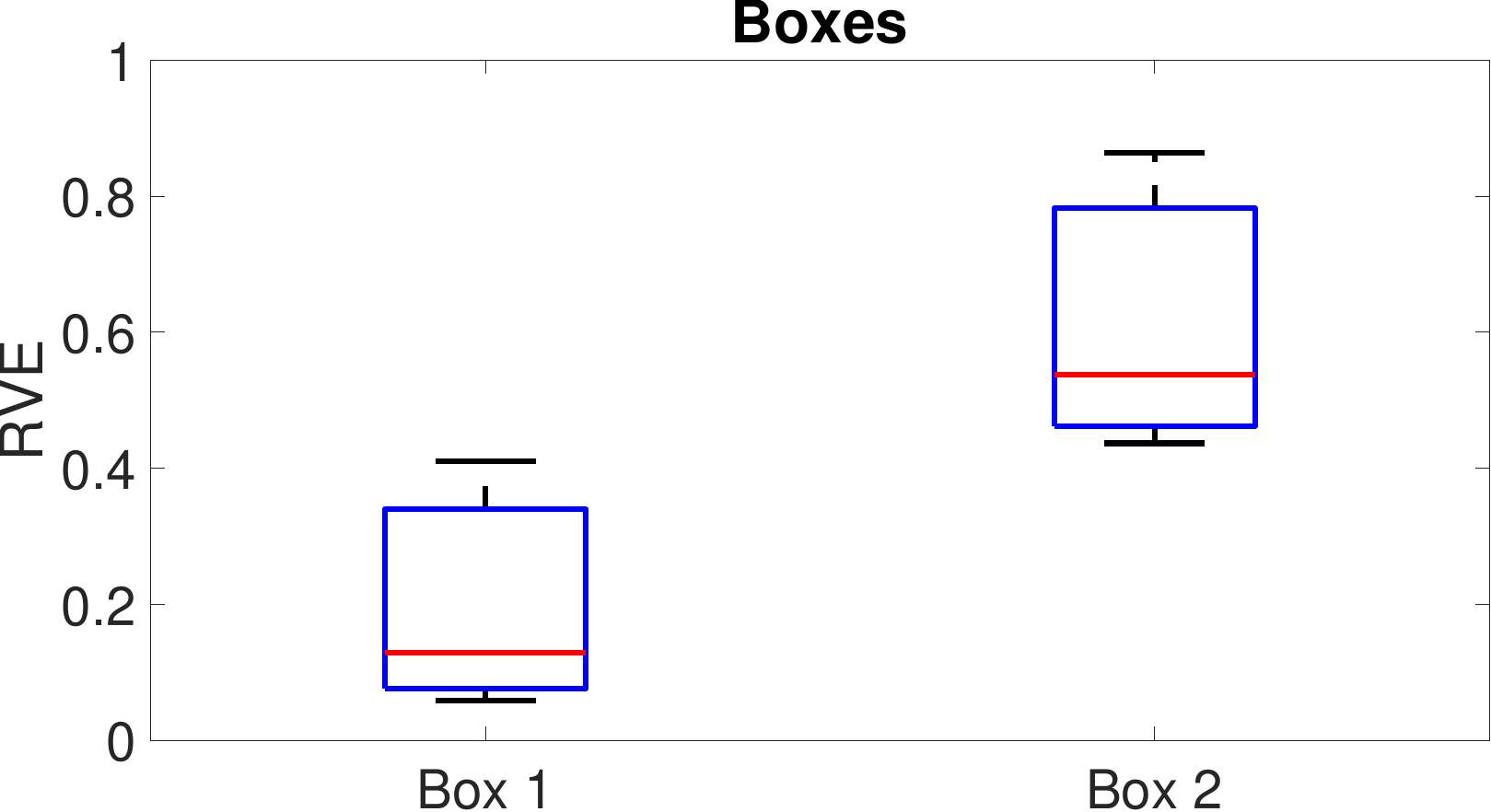}
\caption{Boxes RVE.}
\label{fig:Boxes}
\end{figure}

Figure \ref{fig:Boxes} shows the histograms of the RVE between the estimated and real volumes of the boxes.

\subsection{3D Human Body Model registration results}

\begin{figure}[h]
\centering
\includegraphics[width=0.99\linewidth]{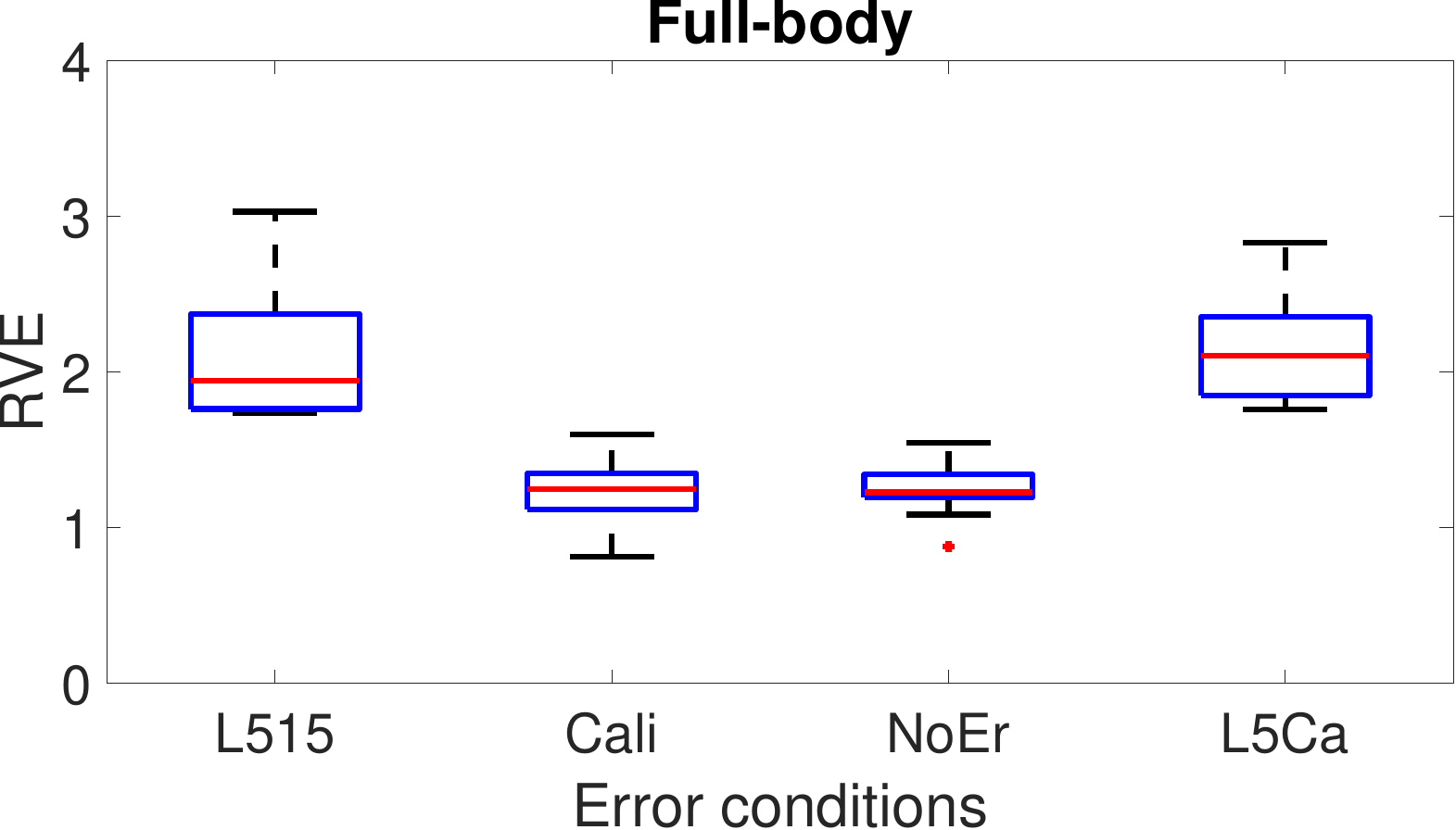}
\caption{Full-body RVE in the four error conditions: L515, Cali, NoEr, and L5Ca}
\label{fig:FullbodyVolume}
\end{figure}

Table \ref{tab:Results} presents the mean and standard deviation of the RVE of the full-body and the segment under the four different error conditions.
\begin{table}[h]
    \caption{Mean and standard deviation (Std Dev) of the full-body and segments RVE in the four different error conditions: Cali, L515, L5Ca, and NoEr.}
    \centering
    \begin{tabular}{|l|l|l|l|}
        \hline
         \textbf{Segment}  & \textbf{Error} & \textbf{Mean [\%]} & \textbf{Std Dev [\%]} \\
         \hline
         Full-body & Cali & 1.23 & 0.21 \\
                   & L515 & 2.14 & 0.47 \\
                   & L5Ca & 2.13 & 0.32 \\
                   & NoEr & 1.23 & 0.18 \\ 
         \hline
         Torso & Cali & 5.50 & 3.08 \\
               & L515 & 6.27 & 2.65 \\
               & L5Ca & 6.72 & 3.41 \\
               & NoEr & 5.07 & 3.58 \\
         \hline
         Left Arm & Cali & 10.63 & 4.93 \\
                  & L515 & 13.47 & 7.69 \\
                  & L5Ca & 14.30 & 6.80 \\
                  & NoEr & 5.82  & 4.28 \\
         \hline
         Right Arm & Cali & 10.75 & 6.00 \\
                   & L515 & 12.70 & 7.94 \\
                   & L5Ca & 19.26 & 12.62 \\
                   & NoEr & 10.06 & 6.99 \\
         \hline
         Left Forearm & Cali & 5.69 & 3.53 \\
                      & L515 & 7.77 & 2.49 \\
                      & L5Ca & 5.71 & 3.14 \\
                      & NoEr & 6.21 & 1.74 \\
         \hline
         Right Forearm & Cali & 4.00 & 1.79 \\
                       & L515 & 6.34 & 2.21 \\
                       & L5Ca & 6.99 & 4.40 \\
                       & NoEr & 4.92 & 1.03 \\
         \hline
         Left Tight & Cali & 4.16 & 3.48 \\
                    & L515 & 6.81 & 3.58 \\
                    & L5Ca & 6.27 & 4.34 \\
                    & NoEr & 5.90 & 3.27 \\
         \hline
         Right Tight & Cali & 7.57  & 5.16 \\
                     & L515 & 10.55 & 7.50 \\
                     & L5Ca & 5.43  & 2.44 \\
                     & NoEr & 6.82  & 4.47 \\
         \hline
         Left Shin & Cali & 6.68 & 1.69 \\
                   & L515 & 9.33 & 2.45 \\
                   & L5Ca & 8.48 & 2.59 \\
                & NoEr & 7.13 & 0.87 \\
         \hline
         Right Shin & Cali & 7.61 & 1.49 \\
                    & L515 & 9.75 & 2.13 \\
                    & L5Ca & 9.40 & 1.18 \\
                    & NoEr & 8.33 & 1.44 \\
         \hline
    \end{tabular}
    \label{tab:Results}
\end{table}
For the full-body RVE, we also report histograms to show the effect on the RVE of including different types of errors (Figure \ref{fig:FullbodyVolume}).

Tables \ref{tab:P2PResults} and \ref{tab:P2PResultsSynt} report the comparison between the results achieved by the Point2PartVolume (P2PV) and BSV methods. 
Table \ref{tab:P2PResults} shows the comparison between the average volume prediction accuracies achieved by P2PV with the two BUFF subjects (P2PV - BUFF) and BSV in the NoEr condition (BSV - NoEr), and those achieved by P2PV with the three PDT13 subjects and BSV in the L5Ca condition (BSV - L5Ca). In the first case, we considered the NoEr condition since the BUFF dataset \cite{zhang2017detailed} is created employing a 3D multi-camera scanner system. In the latter, the L5Ca condition must be considered because the PDT13 dataset is acquired with a Microsoft Kinect, which has low resolution and calibration problems.
Table \ref{tab:P2PResultsSynt} shows the comparison between the percentage of the sample with accuracy over the $90\%$ accuracy threshold, as reported by Hu et al. \cite{hu2023point2partvolume}, between the unseen synthetic datasets (P2PV - Synt), which can be considered error-free since the P2PV is trained with synthetic data, and the BSV - NoEr condition. 

%Table \ref{tab:P2PResults} reports the comparison between the average volume prediction accuracies achieved by the Point2PartVolume (P2PV) method and our BSV pipeline. In particular, the P2PV results obtained with the two BUFF subjects (P2PV - BUFF) and the unseen synthetic datasets (P2PV - Synt) are compared to the NoEr condition of the BSV pipeline (BSV - NoEr) because both datasets contain highly accurate data. Zhang et al. \cite{zhang2017detailed} use a multi-camera scanner system and the synthetic data can be considered error-free since the P2PV is trained with synthetic data. However, it is important to note that the volume prediction accuracy reported for the P2Pv - Synt case is the accuracy of the $90\%$ confidence level. The P2PV results obtained with the two PDT13 subjects (P2PV - PDT13) must be compared to the L5Ca error condition (BSV - L5Ca), since the PDT13 dataset is acquired with a Microsoft Kinect, which has low resolution and calibration problems. 
%achieved an average volume prediction accuracy of $55.48\%$ with unseen synthetic data, $92.82\%$ with two BUFF subjects, and $64.79\%$ with three PDT13 subjects.
\begin{table*}[t]
\centering
    \caption{Comparison between volume prediction accuracies of P2PV -BUFF and BSV - NoEr, and between P2PV - PDT13 and BSV -L5Ca}    
    \begin{tabular}{|l|a|a|f|f|}
        \hline
         \textbf{Segment}  & \textbf{P2PV - BUFF} & \textbf{BSV - NoEr} & \textbf{P2PV - PDT13} & \textbf{BSV - L5Ca} \\
         \hline         
         Torso          & 88.64\% &  94.93\% & 87.41\% & 93.30\% \\
         Left Full Arm  & 96.97\% &  95.49\% & 28.09\% & 90.47\% \\
         Right Full Arm & 98.38\% &  94.15\% & 29.98\% & 88.77\% \\
         Left Full Leg  & 93.53\% &  93.40\% & 69.31\% & 94.40\% \\
         Right Full Leg & 89.88\% &  94.63\% & 61.01\% & 94.08\% \\
         Full-body      & 89.26\% &  98.77\% & 84.39\% & 97.87\% \\
         \hline
    \end{tabular}
    \label{tab:P2PResults}
\end{table*}

\begin{table}[t]
\centering
    \caption{Comparison between volume prediction accuracies of P2PV - Synt and BSV - NoEr}    
    \begin{tabular}{|l|l|l|}
        \hline
         \textbf{Segment}  & \textbf{P2PV - Synt} & \textbf{BSV - NoEr} \\
         \hline         
         Torso          & 94.4\% & 90\% \\
         Left Full Arm  & 53.6\% & 100\% \\
         Right Full Arm & 64.4\% & 70\% \\
         Left Full Leg  & 54.7\% & 70\% \\
         Right Full Leg & 42.2\% & 80\% \\
         Full-body      &  -     & 100\% \\
         \hline
    \end{tabular}
    \label{tab:P2PResultsSynt}
\end{table}

\subsection{Real person reconstruction results}

\begin{figure}[t]
     \centering
     \begin{subfigure}[h]{0.49\columnwidth}
        \centering
         \includegraphics[width=1in]{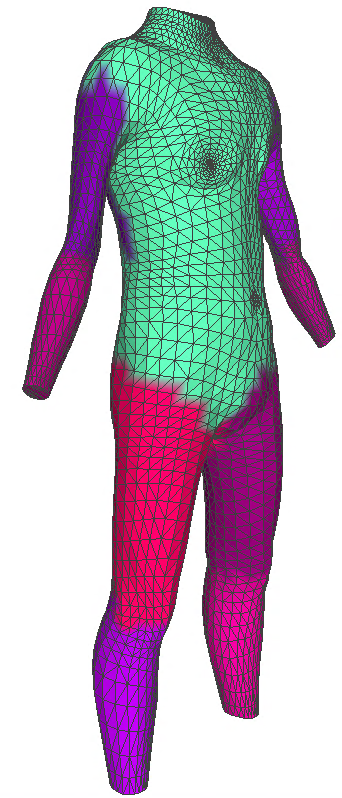}
         \caption{Male subject.}
         \label{fig:AleFit}
     \end{subfigure}
     \hfill
     \begin{subfigure}[h]{0.49\columnwidth}
         \includegraphics[width=0.61\columnwidth]{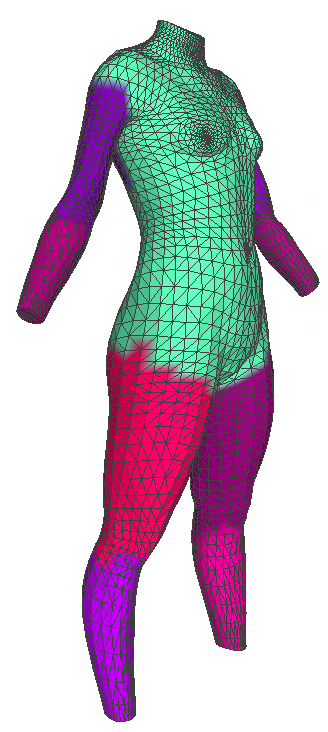}
         \caption{Female subject.}
         \label{fig:GiuliaFit}
     \end{subfigure}
    \caption{Segmented fitted mesh obtained with the male (a) and female (b) subject.}
    \label{fig:BSVfit}
\end{figure}

Figure \ref{fig:BSVfit} shows the segmented fitted meshes of the two real acquisitions on a male (Figure \ref{fig:AleFit}) and a female (Figure \ref{fig:GiuliaFit}) subject obtained with BSV. The estimated volumes evaluated on the fitted meshes are $0.078m^3$ and $0.058m^3$, respectively. Thus, the estimated masses are $78Kg$ and $58Kg$, and, considering that the real masses are $75Kg$ and $60Kg$, the RME is $4\%$ and $3.3\%$.

\section{Discussion} \label{sec:Discussion}

The results presented in the previous Section showed that the BSV estimation pipeline is a valid method to estimate full-body and segment volumes.
The L515 RGB-D camera has high repeatability and precision of calibration (Table \ref{tab:CalEr}). It has low standard deviation and average values. In particular, the rotation errors are lower than $1 deg$ and the translation error are less than $1 cm$ on the vertical and transversal axes and slightly bigger on the longitudinal axis.
%The mean and standard deviations reported in Table \ref{tab:CalEr} show that the L515 RGB-D camera has high repeatability and precision of calibration. 
However, since these errors are constant along the different acquisitions, after each system setup, the calibration offsets can be removed to improve the volume prediction accuracy.

The RVEs calculated with the two boxes are less than $1\%$ (Figure \ref{fig:Boxes}) and prove the ability of the 3D registration algorithm to reconstruct consistent meshes of objects with simple shapes starting from an incomplete PC due to the lack of lateral views.

When dealing with 3D human body models, the average full-body RVE is on the same level, between $1\%$ and $2\%$, thus proving that the 3D registration method can accurately reconstruct a non-rigid shape as the human body starting from just front and back views.
In addition, the calibration errors do not negatively influence the RVE (Figure \ref{fig:FullbodyVolume}), and the L515 noise error causes a limited growth of the RVE ($0.91\%$ for the prediction of the whole body volume (Table \ref{tab:Results})).
The RVEs obtained with the segment volumes are slightly higher, but generally lower $10\%$, except for the upper part of the arms, which suffer from some segmentation irregularities.  

Tables \ref{tab:P2PResults} and \ref{tab:P2PResultsSynt} show that BSV reached higher accuracies than those obtained by the P2PV method, except for a few segments when the P2PV is tested on the BUFF subjects. However, the P2PV method considered only $2$ BUFF subjects and thus has a lower statistical validity than our results, which are the average values computed on the $10$ FAUST subjects.
In addition, the results presented in Table \ref{tab:P2PResultsSynt} prove that their method is not very generalizable since, even if they trained it on synthetic data, just slightly more than $50\%$ of the segments have accuracies higher than $90\%$, except for the torso, when they tested on unseen synthetic data.
%In addition, the results presented in Table \ref{tab:P2PResultsSynt} prove that their method is not very generalizable since, even if they trained it on synthetic data, when they tested on unseen synthetic data, except for the torso, just over $50\%$ of the other segments have accuracies higher than $90\%$.

\section{Conclusion} \label{sec:Conclusion}

The proposed BSV estimation pipeline fills the SoA lack of a low-cost 3D camera system able to estimate the body mass distribution with high accuracy, especially for health applications, sport biomechanics, and ergonomic assessment.
The system combined a minimum number of RGB-D cameras and a new non-rigid registration technique in order to provide a detailed 3D human body model with a limited system complexity. 
The volume accuracy of both whole body and body parts is higher than that of other systems, which use one or two RGB-D cameras. In particular, we compared our results to a method at the SoA and we showed the superiority of our estimation pipeline.
In addition, we segmented the limbs to be able to evaluate the ratio between the proximal and distal parts of the body.
The BSV estimation pipeline can also be used to compute the BVI with simplified BMI-like formulas. However, future developments include further segmentation of the torso into chest, abdomen, and pelvis, and the estimation of other anthropometric measurements, such as waist girth, in order to get an exhaustive system that can also estimate the BVI with higher accuracy.

\bibliographystyle{unsrt}
\bibliography{Bibliography}

\begin{thebibliography}{10}

\bibitem{keys2014indices}
A.~Keys, F.~Fidanza, M.~J. Karvonen, N.~Kimura, and H.~L. Taylor.
\newblock Indices of relative weight and obesity.
\newblock {\em IJEPBF}, 43(3):655--665, 2014.

\bibitem{piche2018overview}
M.~Pich{\'e}, P.~Poirier, I.~Lemieux, and J.~Despr{\'e}s.
\newblock Overview of epidemiology and contribution of obesity and body fat distribution to cardiovascular disease: an update.
\newblock {\em PCD}, 61(2):103--113, 2018.

\bibitem{payton2007biomechanical}
C.~J. Payton and R.~Bartlett.
\newblock {\em Biomechanical evaluation of movement in sport and exercise}.
\newblock Routledge Abingdon, Oxon, UK, 2007.

\bibitem{lloyd2024future}
D.~Lloyd.
\newblock The future of in-field sports biomechanics: Wearables plus modelling compute real-time in vivo tissue loading to prevent and repair musculoskeletal injuries.
\newblock {\em Sports Biomech.}, 23(10):1284--1312, 2024.

\bibitem{stefanyshyn2015biomechanics}
D.~J. Stefanyshyn and J.~W. Wannop.
\newblock Biomechanics research and sport equipment development.
\newblock {\em Sports Eng.}, 18(4):191--202, 2015.

\bibitem{stevenson2004suite}
J.~Stevenson and et~al.
\newblock A suite of objective biomechanical measurement tools for personal load carriage system assessment.
\newblock {\em Ergonomics}, 47(11):1160--1179, 2004.

\bibitem{katch1967estimation}
F.~Katch, E.~D. Michael, and S.~M. Horvath.
\newblock Estimation of body volume by underwater weighing: description of a simple method.
\newblock {\em J. Appl. Physiol.}, 23(5):811--813, 1967.

\bibitem{seidell1990imaging}
J.~C. Seidell, C.~Bakker, and K.~van~der Kooy.
\newblock Imaging techniques for measuring adipose-tissue distribution--a comparison between computed tomography and 1.5-t magnetic resonance.
\newblock {\em AJCN}, 51(6):953--957, 1990.

\bibitem{bartol2021review}
K.~Bartol, David Bojani{\'c}, Tomislav Petkovi{\'c}, and Tomislav Pribani{\'c}.
\newblock A review of body measurement using 3d scanning.
\newblock {\em IEEE Access}, 9:67281--67301, 2021.

\bibitem{daanen20133d}
H.~A. Daanen and F.~B. Ter~Haar.
\newblock 3d whole body scanners revisited.
\newblock {\em Displays}, 34(4):270--275, 2013.

\bibitem{loper2023smpl}
M.~Loper, N.~Mahmood, J.~Romero, G.~Pons-Moll, and M.~J. Black.
\newblock Smpl: A skinned multi-person linear model.
\newblock In {\em Seminal Graphics Papers: Pushing the Boundaries, Volume 2}, pages 851--866, 2023.

\bibitem{anguelov2005scape}
D.~Anguelov and et~al.
\newblock Scape: shape completion and animation of people.
\newblock In {\em ACM Siggraph 2005 Papers}, pages 408--416, 2005.

\bibitem{caesar2002}
CAESAR website.
\newblock [online]. available: https://humanshape.org/caesar/.
\newblock 2002.

\bibitem{bogo2014faust}
F.~Bogo, J.~Romero, M.~Loper, and M.~J. Black.
\newblock Faust: Dataset and evaluation for 3d mesh registration.
\newblock In {\em IEEE CVPR}, pages 3794--3801, 2014.

\bibitem{kwok2014volumetric}
T.~Kwok, K.~Yeung, and C.~C. Wang.
\newblock Volumetric template fitting for human body reconstruction from incomplete data.
\newblock {\em J. Manuf. Syst.}, 33(4):678--689, 2014.

\bibitem{cui2013kinectavatar}
Y.~Cui, W.~Chang, Tobias N{\"o}ll, and Didier Stricker.
\newblock Kinectavatar: fully automatic body capture using a single kinect.
\newblock In {\em ACCV 2012}, pages 133--147. Springer, 2013.

\bibitem{tong2012scanning}
J.~Tong, J.~Zhou, L.~Liu, Z.~Pan, and H.~Yan.
\newblock Scanning 3d full human bodies using kinects.
\newblock {\em IEEE TVCG}, 18(4):643--650, 2012.

\bibitem{cheng2018registration}
L.~Cheng and et~al.
\newblock Registration of laser scanning point clouds: A review.
\newblock {\em MDPI Sensors}, 18(5):1641, 2018.

\bibitem{wang2024real}
H.~Wang, F.~Lai, and F.~Wang.
\newblock Real-time multiple human height measurements with occlusion handling using lidar and camera of a mobile device.
\newblock {\em IEEE Access}, 2024.

\bibitem{oberhofer2024feasibility}
Katja Oberhofer, C{\'e}line Knopfli, Basil Achermann, and Silvio~R Lorenzetti.
\newblock Feasibility of using laser imaging detection and ranging technology for contactless 3d body scanning and anthropometric assessment of athletes.
\newblock {\em Sports}, 12(4):92, 2024.

\bibitem{su2020robustfusion}
Z.~Su and et~al.
\newblock Robustfusion: Human volumetric capture with data-driven visual cues using a rgbd camera.
\newblock In {\em ECCV 2020}, pages 246--264. Springer, 2020.

\bibitem{liu2016template}
Z.~Liu and et~al.
\newblock Template deformation-based 3-d reconstruction of full human body scans from low-cost depth cameras.
\newblock {\em Trans Cybern.}, 47(3):695--708, 2016.

\bibitem{deng2022survey}
B.~Deng, Y.~Yao, R.~M. Dyke, and J.~Zhang.
\newblock A survey of non-rigid 3d registration.
\newblock In {\em Computer Graphics Forum}, pages 559--589. Wiley Online Library, 2022.

\bibitem{amberg2007optimal}
B.~Amberg, S.~Romdhani, and T.~Vetter.
\newblock Optimal step nonrigid icp algorithms for surface registration.
\newblock In {\em CVPR}, pages 1--8. IEEE, 2007.

\bibitem{li2018articulatedfusion}
C.~Li, Z.~Zhao, and X.~Guo.
\newblock Articulatedfusion: Real-time reconstruction of motion, geometry and segmentation using a single depth camera.
\newblock In {\em ECCV}, pages 317--332, 2018.

\bibitem{allen2003space}
B.~Allen, B.~Curless, and Zoran Popovi{\'c}.
\newblock The space of human body shapes: reconstruction and parameterization from range scans.
\newblock {\em ACM TOG}, 22(3):587--594, 2003.

\bibitem{sussmuth2008reconstructing}
Jochen S{\"u}{\ss}muth, Marco Winter, and G{\"u}nther Greiner.
\newblock Reconstructing animated meshes from time-varying point clouds.
\newblock In {\em Computer Graphics Forum}, pages 1469--1476. Wiley Online Library, 2008.

\bibitem{sorkine2007rigid}
O.~Sorkine and M.~Alexa.
\newblock As-rigid-as-possible surface modeling.
\newblock In {\em SGP}, volume~4, pages 109--116. Citeseer, 2007.

\bibitem{yang2019global}
J.~Yang, D.~Guo, K.~Li, Z.~Wu, and Y.~Lai.
\newblock Global 3d non-rigid registration of deformable objects using a single rgb-d camera.
\newblock {\em IEEE TIP}, 28(10):4746--4761, 2019.

\bibitem{chen2017rigidity}
S.~Chen, L.~Gao, Y.~Lai, and S.~Xia.
\newblock Rigidity controllable as-rigid-as-possible shape deformation.
\newblock {\em Graphical Models}, 91:13--21, 2017.

\bibitem{jiang2017consistent}
T.~Jiang and et~al.
\newblock Consistent as-similar-as-possible non-isometric surface registration.
\newblock {\em The Visual Computer}, 33:891--901, 2017.

\bibitem{wells2000assessment}
J.~Wells, I.~Douros, N.~Fuller, M.~Elia, and L.~Dekker.
\newblock Assessment of body volume using three-dimensional photonic scanning.
\newblock {\em Annals of the New York Academy of Sciences}, 904(1):247--254, 2000.

\bibitem{chiu2018automated}
C.~Chiu, D.~L. Pease, S.~Fawkner, and R.~H. Sanders.
\newblock Automated body volume acquisitions from 3d structured-light scanning.
\newblock {\em CBM}, 101:112--119, 2018.

\bibitem{pirker2009human}
K.~Pirker, Matthias R{\"u}ther, Horst Bischof, Falko Skrabal, and Georg Pichler.
\newblock Human body volume estimation in a clinical environment.
\newblock {\em AAPR/OAGM}, 2009.

\bibitem{pfitzner2015libra3d}
C.~Pfitzner and et~al.
\newblock Libra3d: Body weight estimation for emergency patients in clinical environments with a 3d structured light sensor.
\newblock In {\em ICRA}, pages 2888--2893. IEEE, 2015.

\bibitem{nuzzi2025measurement}
C.~Nuzzi and et~al.
\newblock Measurement of human body segment properties using low-cost rgb-d cameras.
\newblock {\em MDPI Sensors}, 25(5):1515, 2025.

\bibitem{cook2013using}
T.~S. Cook, G.~Couch, T.~J. Couch, W.~Kim, and W.~W. Boonn.
\newblock Using the microsoft kinect for patient size estimation and radiation dose normalization: Proof of concept and initial validation.
\newblock {\em JDI}, 26:657--662, 2013.

\bibitem{he2018volumeter}
Q.~He, Y.~Ji, D.~Zeng, and Z.~Zhang.
\newblock Volumeter: 3d human body parameters measurement with a single kinect.
\newblock {\em IET Computer Vision}, 12(4):553--561, 2018.

\bibitem{hu2023point2partvolume}
P.~Hu and et~al.
\newblock Point2partvolume: Human body volume estimation from a single depth image.
\newblock {\em IEEE Transactions on Instrumentation and Measurement}, 72:1--12, 2023.

\bibitem{zhang2017detailed}
C.~Zhang, S.~Pujades, M.~J. Black, and G.~Pons-Moll.
\newblock Detailed, accurate, human shape estimation from clothed 3d scan sequences.
\newblock In {\em IEEE CVPR}, pages 4191--4200, 2017.

\bibitem{helten2013personalization}
T.~Helten and et~al.
\newblock Personalization and evaluation of a real-time depth-based full body tracker.
\newblock In {\em 3DV}, pages 279--286. IEEE, 2013.

\bibitem{garcia2024development}
Fabi{\'a}n~I. Garc{\'\i}a~F., M.~Kl{\"u}nder, M.~T. L{\'o}pez~Teros, C.~A. Mu{\~n}oz~Iba{\~n}ez, and M.~A. Padilla~Casta{\~n}eda.
\newblock Development and validation of a method of body volume and fat mass estimation using three-dimensional image processing with a mexican sample.
\newblock {\em Nutrients}, 16(3):384, 2024.

\bibitem{wilson2013ratio}
J.~P. Wilson, A.~M. Kanaya, B.~Fan, and J.~A. Shepherd.
\newblock Ratio of trunk to leg volume as a new body shape metric for diabetes and mortality.
\newblock {\em PLoS One}, 8(7):e68716, 2013.

\bibitem{hattori2021upper}
Y.~Hattori, K.~Hayashi, S.~Sakamoto, and K.~Doi.
\newblock Upper extremity volume/total body volume ratio for evaluation of upper extremity lymphedema.
\newblock {\em Annals of Plastic Surgery}, 86(1):35--38, 2021.

\bibitem{bluher2019new}
Matthias Bl{\"u}her and Ulrich Laufs.
\newblock New concepts for body shape-related cardiovascular risk: role of fat distribution and adipose tissue function.
\newblock {\em EHJ}, 40(34):2856--2858, 2019.

\bibitem{lugaresi2019mediapipe}
C.~Lugaresi and et~al.
\newblock Mediapipe: A framework for building perception pipelines.
\newblock {\em arXiv preprint arXiv:1906.08172}, 2019.

\bibitem{bodypix2019}
Bodypix website.
\newblock [online]. available: https://blog.tensorflow.org/2019/11/updated-bodypix-2.html.
\newblock 2019.

\bibitem{li2009robust}
H.~Li, B.~Adams, L.~J. Guibas, and M.~Pauly.
\newblock Robust single-view geometry and motion reconstruction.
\newblock {\em ACM ToG}, 28(5):1--10, 2009.

\bibitem{zell2013elastiface}
E.~Zell and M.~Botsch.
\newblock Elastiface: Matching and blending textured faces.
\newblock In {\em NPAR}, pages 15--24, 2013.

\end{thebibliography}

\begin{IEEEbiography}[{\includegraphics[width=1in,height=1.25in,clip,keepaspectratio]{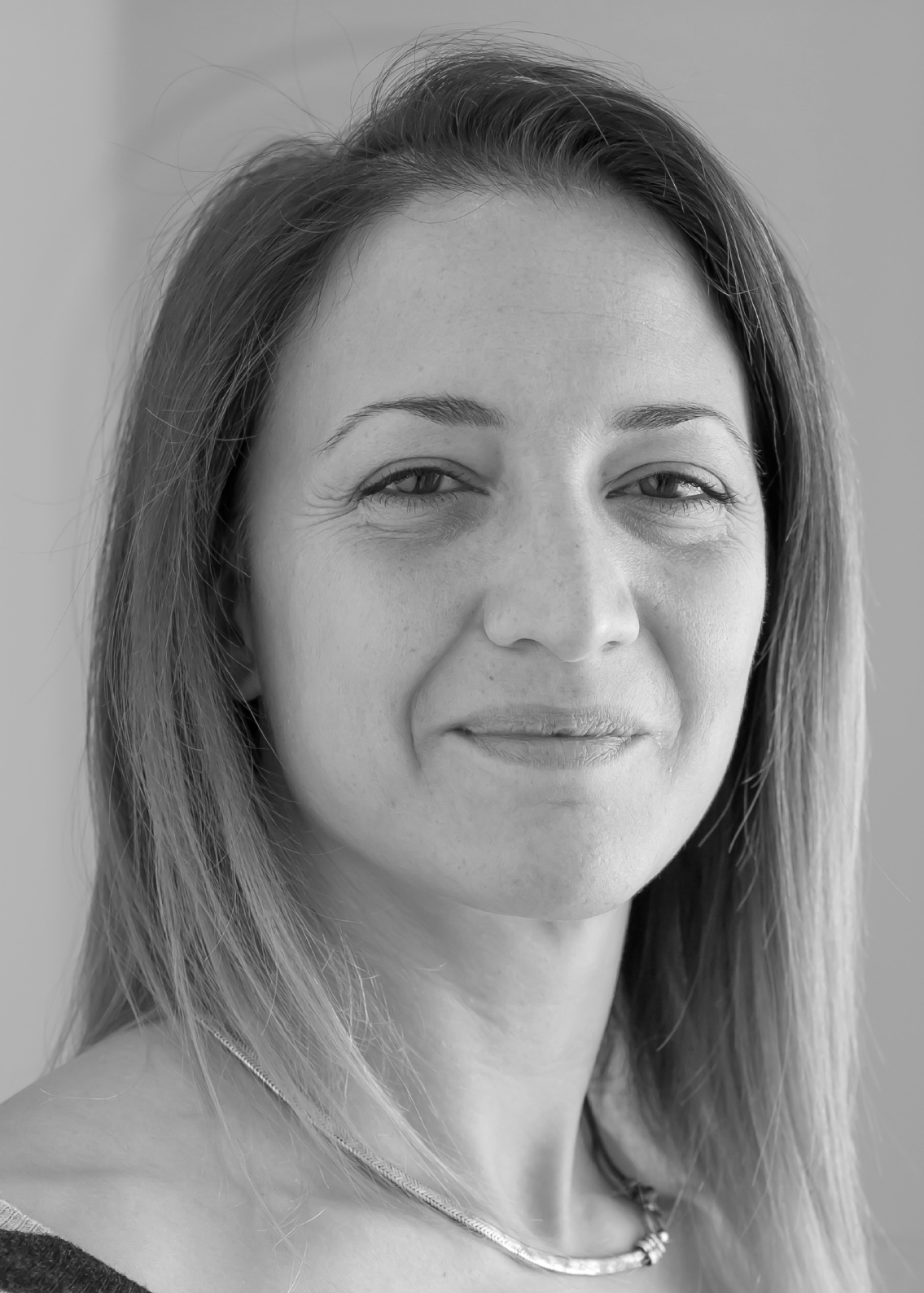}}]{Giulia Bassani} received the M.S. degree in biomedical engineering from the University of Pisa in 2012 and the Ph.D. degree in emerging digital technologies from the Scuola Superiore Sant’Anna (SSSA) in 2017.
She is a Research Fellow at the Mechanical Intelligent Institute, SSSA. 
She participated in national and industrial projects aimed at the automatic assessment of the biomechanical overload risk with a wearable sensor network. She contributed to the Foresight and Technology Injection - H12023 and the EU Exosmooth projects.
Her main research interests include sEMG acquisition and processing, human body motion acquisition and analysis, wearable sensor networks, estimation of ergonomic risk, embedded wearable energy harvesting systems, and deep learning.
\end{IEEEbiography}

\begin{IEEEbiography}[{\includegraphics[width=1in,height=1.25in,clip,keepaspectratio]{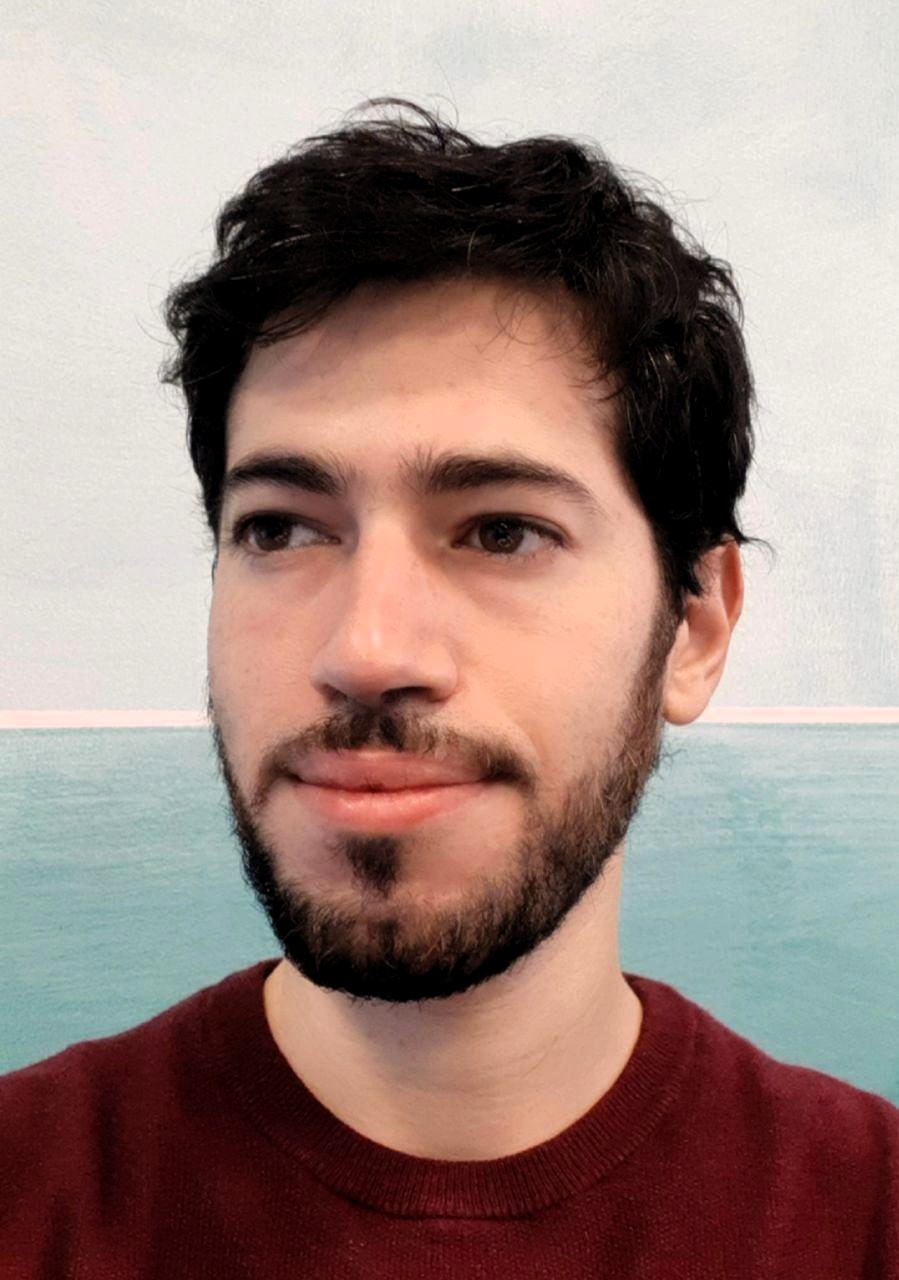}}]{Emilio Maoddi} is a Research Fellow at Leonardo Innovation Labs, Leonardo S.p.A. for the autonomous systems laboratory. After a BSc in Electronics Engineering from Polytechnic University of Turin, he received his MSc in Robotics Engineering from University of Pisa in 2022. From 2022 to 2024, he was a Researcher at Ericsson's Research. 
Current research interests include autonomous systems, focusing on cognitive human machine interfaces, and scalable autonomy.
The research presented herein was conducted while he was enrolled at Scuola Superiore Sant'Anna. Leonardo S.p.A. was not involved in the development of this article or its content.
\end{IEEEbiography}

\begin{IEEEbiography}[{\includegraphics[width=1in,height=1.25in,clip,keepaspectratio]{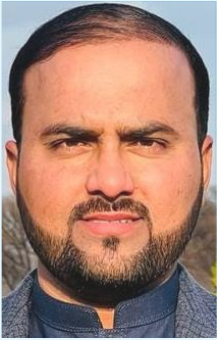}}]{Usman Asghar} received a BSc degree in Computer Science from the University of Gujrat in 2017 and an MSc degree in Computer Science from the University of Engineering and Technology in Lahore in 2021, and is now enrolled in a multidisciplinary PhD program in Scuola Superiore Sant’Anna. His research interests include artificial intelligence, computer vision, and medical image processing.
\end{IEEEbiography}

%\vskip 0pt plus -1fil
\begin{IEEEbiography}[{\includegraphics[width=1in,height=1.25in,clip,keepaspectratio]{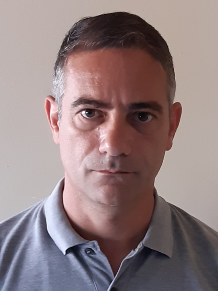}}]{Carlo Alberto Avizzano, Ass. Prof.} received the B.S. degree in Control and Automation from University of Pisa in 1995 and the PhD. in Robotics in 1999. He is currently the Director of the Intelligent Automation System Laboratory, and the Coordinator of the Department of Excellence in Robotics and Artificial Intelligence at Scuola Superiore Sant'Anna, Pisa (IT). Avizzano's research interests include intelligent sensing and control systems, Industry 4.0, smart wearable devices, human-robot interfaces with high cognitive capabilities, autonomous vehicles and drones in cognitive aware high perception tasks. Avizzano's skills include control, robotics, computer vision, artificial intelligence, mechatronics, embedded systems, and haptics. He is cooperating in EU, National, Regional and Industrial projects since 1996. To date, he is owner of about 15 industrial patents and authored more than 170 scientific and peer reviewed papers in journals and conference proceedings.
\end{IEEEbiography}

%\vskip 0pt plus -1fil
\begin{IEEEbiography}[{\includegraphics[width=1in,height=1.25in,clip,keepaspectratio]{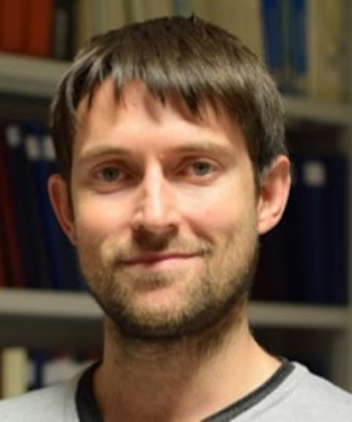}}]{Alessandro Filippeschi}  received M.S. degree in Mechanical Engineering in 2007 from University of Pisa and a PhD in Perceptual Robotics in 2012 from Scuola Superiore Sant'Anna. He is Assistant Professor of Applied Mechanics at Scuola Superiore Sant’Anna. He participated in more than 20 national and EU projects. His research interests include the capture and analysis of human motion, estimation of the ergonomic risks, and the development of exoskeletons and haptic interfaces for human-robot interaction. He is a co-founder of Wearable Robotics S.r.L.
\end{IEEEbiography}

\end{document}